\documentclass{revtex4}
\usepackage{amsmath}
\usepackage{amssymb}
\usepackage{amsfonts}
\usepackage{bm}
\usepackage[dvips]{color}

\newcommand\hl[1]{{\color{blue} #1}}

\begin{document}

\title{Application of Projection Operator Method \\
to Coarse-Grained Dynamics with Transient Potential}

\author{Takashi Uneyama}
\affiliation{%
  JST-PRESTO, and Department of Materials Physics, Graduate School of Engineering,
  Nagoya University, Furo-cho, Chikusa, Nagoya 464-8603,
  Japan
}%

\begin{abstract}
We show that the coarse-grained dynamics model with the time-dependent and fluctuating potential
(transient potential) can be derived from the microscopic Hamiltonian dynamics.
The concept of the transient potential was first introduced rather phenomenologically,
and its relation to the underlying microscopic dynamics has not been clarified yet.
This is in contrast to the generalized Langevin equation, of which 
relation to the microscopic dynamics is well-established.
In this work, we show that the dynamic equations with the transient potential
can be derived for the coupled oscillator model, without any approximations.
It is known that the dynamics of the coupled oscillator model can be
exactly described by the generalized Langevin type equations.
This fact implies that the dynamic equations with the transient potential
can be utilized as a coarse-grained dynamics model in a similar way to the generalized
Langevin equation. Then we show that the dynamic equations for the transient
potential can be also formally derived for the microscopic Hamiltonian dynamics,
without any approximations. We use the projection operator method for the
coarse-grained variables and transient potential. The dynamic equations for
the coarse-grained positions and momenta are similar to those in the Hamiltonian
dynamics, but the interaction potential is replaced by the transient potential.
The dynamic equation for the transient potential is the generalized Langevin
equation with the memory effect.
Our result justifies the use of the transient
potential to describe the coarse-grained dynamics. We propose several
approximations to obtain simplified dynamics model. We show that, under several approximations,
the dynamic equation for the transient potential reduces to the relatively simple Markovian
dynamic equation for the potential parameters.
We also show that with several additional approximations,
the approximate dynamics model further reduces to 
the Markovian Langevin type equations with the transient potential.
\end{abstract}

\maketitle

%

\section{Introduction}
\label{introduction}

Soft matters such as polymers exhibit complex dynamics at relatively long
time scales.
To describe the dynamics of soft matters, mesoscopic
coarse-grained models are useful. For example, we can efficiently
perform simulations for slow relaxation processes of entangled polymers
by utilizing some coarse-grained models\cite{Doi-Edwards-book,Padding-Briels-2011}.
In many cases, coarse-grained models are designed rather phenomenologically.
As a result, the dynamic equations and interaction models in phenomenological coarse-grained
models are not fully justified, and sometimes the connections between 
a coarse-grained model and a microscopic (or atomistic) model are not clear, neither.
Some statistical mechanical methods to relate
microscopic and mesoscopic coarse-grained models are desired.

If we limit ourselves to static structures, the connection between
the coarse-grained and microscopic models can be established systematically.
To obtain the effective interaction potential in a coarse-grained model,
for example, we can employ the iterative Boltzmann inversion method\cite{Reith-Putz-MullerPlathe-2003}.
With the iterative
Boltzmann inversion method, we can construct the accurate effective interaction
potentials starting from the underlying microscopic model.
Then we can efficiently reproduce static structures by the coarse-grained model with
the effective interaction potentials.
Such a coarse-graining procedure can be hierarchically performed and effective interaction
potentials at several different scales can be systematically construct\cite{Harmandaris-Adhikari-vanderVegt-Kremer-2006}.

However, if we want to reproduce the dynamics by
the mesoscopic coarse-grained model,
the situation becomes tough. To reasonably describe the dynamics at the mesoscopic
level, we need some transport coefficients and memory effects
which are consistent with the underlying microscopic dynamics.
The projection operator method\cite{Kawasaki-1973,Grabert-book,Espanol-2004,Hijon-Espanol-VandenEijnden-DelgadoBuscalioni-2010,Dengler-2016} tells us
a straightforward way to calculate the
coarse-grained dynamic equations for mesoscopic coarse-grained variables.
By operating the projection operator to the time evolution of the coarse-grained
variables, we obtain the generalized Langevin equations. The thus obtained
generalized Langevin equations contain so-called the memory kernels which
represent the memory effect due to the eliminated fast degrees of freedom.
With some additional approximations (the Gaussian noise
approximation and the Markov approximation), the dynamic equations reduce to
well-known simple Langevin equations.
The projection operator method has been utilized to derive 
mesoscopic models such as
the time-dependent density functional\cite{Yoshimori-2005},
the dissipative particle dynamics (DPD)\cite{Kinjo-Hyodo-2007},
and hybrid models\cite{Espanol-2009}.

The projection operator method
for the generalized Langevin equation is an established method and widely
accepted as the standard way of coarse-graining. However, it is not a unique
way to obtain the coarse-grained dynamic equations.
We can employ other coarse-graining methods.
For some targets, the generalized Langevin equation would be convenient,
and for some other targets, some other
coarse-graining methods would be convenient.
One simple yet powerful coarse-graining method is the modeling with
the transient force (or the transient potential). The concept of the transient force was
first proposed by Kindt and Briels\cite{Kindt-Briels-2007} in
the responsive particle dynamics (RaPiD) model for entangled polymers,
and then generalized to various soft matter systems\cite{Briels-2009,Padding-Briels-2011}.
In the transient potential type models, interaction potentials are modeled
as time-dependent functions which relax toward stable forms.
The slip-spring type models\cite{Likhtman-2005,Uneyama-2011,Chappa-Morse-Zippelius-Muller-2012,Uneyama-Masubuchi-2012,Uneyama-2019}
for entangled polymers would be interpreted as models with transient potentials.
Also, the dynamically heterogeneous motion of particles in supercooled
liquids can be reasonably described with transient potentials\cite{Hachiya-Uneyama-Kaneko-Akimoto-2019,Uneyama-2020}.
Although the concept of the transient potential seems to be promising, the
dynamics models listed above are constructed phenomenologically.
Whether the use of the transient potentials
is theoretically acceptable or not is not understood.
The connection between the transient potential and the
underlying microscopic model is not clear, neither.
Theories to connect the transient potential and the underlying microscopic
dynamics are demanding.

Recently, the author attempted to derive the transient potential
from the microscopic Langevin dynamics and justify the dynamics models
with the transient potentials.
The Langevin equation with
the transient potential (LETP)\cite{Uneyama-2020} was proposed
as a general coarse-grained dynamics model.
The LETP was obtained by introducing the transient potential
as an additional and
auxiliary degree of freedom to the system, and then formally eliminating
the fast degrees of freedom.
Although the previous work partly justified the use of the transient potential,
it does not give the explicit dynamic equation for the transient potential.
The transient potential obeys the stochastic process with the memory effect,
and we only have the formal path probability for it.
In addition, an overdamped Langevin equation was employed as the microscopic
dynamics model. Thus the microscopic model itself was already
(partly) coarse-grained. Therefore, the relation of the LETP to its
underlying microscopic dynamics model is still not fully clear.

To fully justify the transient potential model and establish the connection between dynamics of
the transient potentials and the underlying microscopic dynamics models,
we may need to start the derivation from a microscopic Hamiltonian dynamics.
One may employ the underdamped Langevin equation or the generalized
Langevin equation. However, these dynamic equations are also already (partly)
coarse-grained. Thus we employ the Hamiltonian dynamics as the microscopic
model.
In this work, first we show that the transient potential model can be
exactly derived for the coupled oscillator model, without any approximations (Sec.~\ref{coupled_oscillator_model}).
The coupled oscillator model is a simple Hamiltonian dynamics model, and
is well known as a simple model from which the generalized Langevin type equations can be derived.
The dynamic equations with the transient potential can be derived in a somewhat similar way to the generalized
Langevin type equations.
This fact encourages us to consider the construction of 
the transient potential model for more general Hamiltonian-based microscopic dynamics.

Then we apply the projection operator method to derive the dynamics
model with the transient potential (Sec.~\ref{theory}).
We introduce the microscopic model and its equilibrium properties
(Secs.~\ref{microscopic_dynamics_model} and \ref{equilibrium_properties}),
and derive the effective dynamic equations for coarse-grained variables
with the transient potential (Sec.~\ref{projection_operators_and_dynamic_equations_for_coarse_grained_variables}).
The effective dynamic equations for the coarse-grained positions and momenta are
shown to be simple canonical type equations.
The dynamic equation for the transient potential can be formally expressed
as the stochastic partial differential equation with the memory effect. It can be interpreted as the generalized Langevin equation for the transient potential.
These dynamic equations are obtained without any approximations and thus
they are exact. Therefore we can formally justify the use of the transient
potential to describe coarse-grained mesoscopic dynamics. However, the
exact dynamic equation for the transient potential is just formal and far from tractable. Therefore
we introduce some approximations to obtain simple 
approximate dynamic equation for the transient potential (Secs.~\ref{markov_approximation} and \ref{approximation_by_potential_parameters}).
We show that the dynamic equation for the transient potential can be approximated
as the simplified dynamic equation for the potential parameters.
Therefore, the use of the transient potentials in
phenomenological dynamics dynamics models can be justified.
Finally we discuss the properties of the obtained dynamics model and how we can utilize them
for practical purposes (Sec.~\ref{discussions}).
We propose a possible method to estimate the potential parameter from a given transient potential (Sec.~\ref{estimating_potential_parameters_from_transient_potential}).
We compare our model with the generalized Langevin equation by the standard coarse-graining method (Sec.~\ref{comparison_with_generalized_langevin_equation}).
We show that our dynamic equations reduce to the LETP in the previous work\cite{Uneyama-2020},
by introducing some additional approximations (Sec.~\ref{comparison_with_langevin_equation_with_transient_potential}).

\section{Coupled Oscillator Model}
\label{coupled_oscillator_model}

\subsection{Dynamics Model}
\label{dynamics_model}

The concept of the transient potential was first introduced rather
phenomenologically\cite{Kindt-Briels-2007,Briels-2009,Padding-Briels-2011}.
If we can rewrite microscopic dynamic equations into the dynamic equations with a transient
potential without approximations, the use of the transient potential would
be justified. For such a purpose, a simple and analytically solvable model
would be useful.
Here we consider an analytically solvable toy model: the coupled oscillator model.
The dynamic equations for the coupled oscillator model can be exactly rewritten as the generalized Langevin type equations
in a simple and straightforward way\cite{Ford-Kac-Mazur-1965,Zwanzig-1973,Sekimoto-book}.
This demonstrates the validity of the generalized Langevin equation.
In this section, we show that the coupled oscillator model can also be rewritten
as the dynamic equations with the transient potential.

We consider a simple coupled oscillator model in a one dimensional space,
which consists of $2 N$ degrees of freedom.
We consider a single particle of which
position and momentum at time $t$ are expressed as $Q(t)$ and $P(t)$. The system consists of
this target particle and $(N - 1)$ particles connected to the target particle
by harmonic springs.
We interpret the $(N - 1)$ particles except the target particle
as a heat bath, and call these particles as the bath particles.
We express
the position and momentum of the $j$-th bath particle as $\theta_{j}(t)$
and $\pi_{j}(t)$. Then the degrees of freedom for the heat bath can
be expressed as the $(N - 1)$-dimensional position and momentum vectors,
$\bm{\theta}(t) \equiv [\theta_{1}(t),\theta_{2}(t),\dotsb,\theta_{N - 1}(t)]$
and $\bm{\pi}(t) \equiv [\pi_{1}(t),\pi_{2}(t),\dotsb,\pi_{N - 1}(t)]$.

We employ the following Hamiltonian:
\begin{align}
 \label{hamiltonian_coupled_oscillator}
 \mathcal{H}(Q,P,\bm{\theta},\bm{\pi}) & = \frac{ P^{2}}{2 M}
  + \sum_{j = 1}^{N - 1}
 \frac{\pi_{j}^{2} }{2 \mu_{j}} + U(Q,\bm{\theta}), \\
 \label{potential_coupled_oscillator}
 U(Q,\bm{\theta}) & =  \sum_{j = 1}^{N - 1}
 \frac{1}{2} \kappa_{j} (Q - \theta_{j})^{2},
\end{align}
where $M$ and $\mu_{j}$ represents the mass for the target particle and
the $j$-th bath particle, $U(Q,\bm{\theta})$ is the interaction potential
between the target particle and the heat bath, and $\kappa_{j}$ is the spring constant for the $j$-th bath particle.
From the Hamiltonian \eqref{hamiltonian_coupled_oscillator}, the canonical equations
are given as
\begin{align}
 \label{canonical_equation_coupled_oscillator_q_p}
 \frac{dQ(t)}{dt} & = \frac{P(t)}{M} ,
 & \frac{dP(t)}{dt} & = - \sum_{j} \kappa_{j} (Q(t) - \theta_{j}(t)), \\
 \label{canonical_equation_coupled_oscillator_theta_pi}
 \frac{d\theta_{j}(t)}{dt} & = \frac{\pi_{j}(t)}{\mu_{j}} ,
 & \frac{d\pi_{j}(t)}{dt}  & = - \kappa_{j} (\theta_{j}(t) - Q(t)). 
\end{align}
Eqs~\eqref{canonical_equation_coupled_oscillator_q_p} and \eqref{canonical_equation_coupled_oscillator_theta_pi}
are the dynamic equations of the coupled oscillator model. Because they
are linear differential equations, we can of course solve them straightforwardly.
We consider the situation where we can observe only the target particle, $Q(t)$ and $P(t)$.
Then we will observe that $Q(t)$ and $P(t)$ obey the effective dynamic equations.
In Secs.~\ref{generalized_langevin_equation} and \ref{dynamic_equations_with_transient_potential}, we derive two sets of effective dynamic equations.

\subsection{Generalized Langevin Equation}
\label{generalized_langevin_equation}

By eliminating $\bm{\theta}(t)$ and $\bm{\pi}(t)$ from
eqs~\eqref{canonical_equation_coupled_oscillator_q_p} and \eqref{canonical_equation_coupled_oscillator_theta_pi},
we can construct the effective dynamic equations for the target particle.
Here we briefly show the derivation of the generalized Langevin type equations\cite{Ford-Kac-Mazur-1965,Zwanzig-1973,Sekimoto-book}.
The first equation in eq~\eqref{canonical_equation_coupled_oscillator_q_p} is expressed in terms of $Q(t)$ and $P(t)$
and thus no further treatment is required.
The second equation in eq~\eqref{canonical_equation_coupled_oscillator_q_p}
contains $\bm{\theta}(t)$ and thus we should eliminate it.
Following the standard procedure, we solve eq~\eqref{canonical_equation_coupled_oscillator_theta_pi} as
\begin{equation}
 \label{canonical_equation_coupled_oscillator_theta_formal_solution}
 \theta_{j}(t) = Q(t)
  + (\theta_{j}(0) - Q(0)) \cos(\omega_{j} t) 
  + \frac{\pi_{j}(0)}{\omega_{j} \mu_{j}} \sin(\omega_{j} t)
  - \int_{0}^{t} dt' \, \cos(\omega_{j} (t - t')) \frac{P(t') }{M} ,
\end{equation}
where $\omega_{j} \equiv \sqrt{\kappa_{j} / \mu_{j}}$.

By substituting
eq~\eqref{canonical_equation_coupled_oscillator_theta_formal_solution}
into eq~\eqref{canonical_equation_coupled_oscillator_q_p} and
rearranging terms, we have the following generalized Langevin type equations\cite{Sekimoto-book}:
\begin{align}
 \label{gle_coupled_oscillator_q}
 \frac{dQ(t)}{dt} & = \frac{P(t)}{M} , \\
 \label{gle_coupled_oscillator_p}
 \frac{dP(t)}{dt} & = - \int_{0}^{t} dt' K'(t - t') \frac{1}{M} P(t')
 + \xi'(t) .
\end{align}
Here, $K'(t)$ and $\xi'(t)$
correspond to the memory kernel and the noise, respectively.
They are defined as
\begin{align}
 \label{gle_coupled_oscillator_memory_kernel}
 K'(t) & \equiv \sum_{j} \kappa_{j} \cos(\omega_{j} t), \\
 \label{gle_coupled_oscillator_noise}
 \xi'(t) & \equiv  (\theta_{j}(0) - Q(0)) \cos(\omega_{j} t) 
  + \frac{\pi_{j}(0)}{\omega_{j} \mu_{j}} \sin(\omega_{j} t).
\end{align}
If we can observe only $Q(t)$ and $P(t)$ and cannot observe $\bm{\theta}(t)$ and $\bm{\pi}(t)$, we should interpret $\bm{\theta}(0)$ and $\bm{\pi}(0)$
as random variables. If we assume that these random variables obey the equilibrium
distribution, we have the fluctuation-dissipation relation:
\begin{equation}
 \label{fluctuation_dissipation_relation_gle_coupled_oscillator}
 \langle \xi'(t) \rangle_{\text{eq},0} = 0, \qquad
 \langle  \xi'(t) \xi'(t') \rangle_{\text{eq},0} 
 = k_{B} T K'(|t - t'|).
\end{equation}
Here, $\langle \dots \rangle_{\text{eq},0}$ represents the equilibrium
statistical average over $\bm{\theta}(0)$ and $\bm{\pi}(0)$,
$k_{B}$ is the Boltzmann constant, and $T$ is the temperature.
The equilibrium statistical average is taken for the canonical
probability distribution determined by the Hamiltonian \eqref{hamiltonian_coupled_oscillator}.

Thus we find that the generalized Langevin type equations are derived from the canonical
equations, only with straightforward calculations. The derivation shown
above supports the use of the generalized Langevin equation with the memory
kernel to describe the coarse-grained dynamics.

\subsection{Dynamic Equations with Transient Potential}
\label{dynamic_equations_with_transient_potential}

The generalized Langevin type equations in Sec.~\ref{generalized_langevin_equation}
is not the unique model to describe the coarse-grained dynamics.
We expect that eqs~\eqref{canonical_equation_coupled_oscillator_q_p} and
\eqref{canonical_equation_coupled_oscillator_theta_pi} can be rewritten 
in other forms. Now we consider to rewrite them by utilizing the time-dependent
and fluctuating transient potential in a heuristic way.
If we concentrate only on $Q(t)$ and $P(t)$, the interaction potential
$U(Q(t),\bm{\theta}(t))$ would be replaced by an effective potential for $Q(t)$.
This effective potential will change as time evolves.
Here we may call such an effective potential as the transient potential.
We hypothetically rewrite the potential as $U(Q(t),\bm{\theta}(t)) = \Phi(Q(t),A(t))$,
with
\begin{equation}
 \label{transient_potential_coupled_oscillator}
  \Phi(Q,A) = \frac{1}{2} \kappa_{\text{eff}} (Q - A)^{2} ,
\end{equation}
where $\kappa_{\text{eff}}$ represents the effective spring constant and $A$ is the
effective position of the potential center. The effective position can change in time, and thus we interpret $A(t)$ at time $t$ as
an additional, auxiliary degree of freedom. From eqs~\eqref{transient_potential_coupled_oscillator}
and \eqref{potential_coupled_oscillator},
it is straightforward to show that
the following relations hold:
\begin{equation}
 \label{potential_parameters_coupled_oscillator}
  \kappa_{\text{eff}} = \sum_{j} \kappa_{j}, \qquad A(t) 
  = \frac{1}{\kappa_{\text{eff}}} \sum_{j} \kappa_{j} \theta_{j}(t).
\end{equation}
With the transient potential~\eqref{transient_potential_coupled_oscillator},
the second equation in eq~\eqref{canonical_equation_coupled_oscillator_q_p} can be rewritten as
\begin{equation}
 \label{canonical_equation_coupled_oscillator_p_modified}
 \frac{dP(t)}{dt} 
 =  - \frac{\partial \Phi(Q(t),A(t))}{\partial A(t)} .
\end{equation}

We need to construct the dynamic equation for $A(t)$ to obtain 
the dynamic equations in a closed form. From eqs~\eqref{potential_parameters_coupled_oscillator}
and \eqref{canonical_equation_coupled_oscillator_theta_pi},
we have the following dynamic equation for $A(t)$:
\begin{equation}
 \label{dynamic_equation_coupled_oscillator_a}
 \frac{dA(t)}{dt} = \frac{1}{\kappa_{\text{eff}}} \sum_{j} \kappa_{j} \frac{d\theta_{j}(t)}{dt}
  =  \frac{1}{\kappa_{\text{eff}}} \sum_{j} \omega_{j}^{2} \pi_{j}(t) .
\end{equation}
As the case in Sec.~\ref{generalized_langevin_equation}, we solve eq~\eqref{canonical_equation_coupled_oscillator_theta_pi}
to eliminate $\bm{\pi}(t)$ in
eq~\eqref{dynamic_equation_coupled_oscillator_a}:
\begin{equation}
 \label{canonical_equation_coupled_oscillator_pi_formal_solution}
\begin{split}
  \pi_{j}(t) & = - \omega_{j} \mu_{j} 
  (\theta_{j}(0) - Q(0)) \sin(\omega_{j} t) 
  + \pi_{j}(0) \cos(\omega_{j} t) \\
 & \qquad + \int_{0}^{t} dt' \, 
  \left[ \omega_{j} \mu_{j} \sin(\omega_{j} (t - t')) \frac{dA(t')}{dt'}
   + \cos(\omega_{j}(t - t')) \kappa_{j} (Q(t) - A(t))
   \right].
\end{split}
\end{equation}
By substituting eq~\eqref{canonical_equation_coupled_oscillator_pi_formal_solution} into
eq~\eqref{dynamic_equation_coupled_oscillator_a}, we have the dynamic equation
for $A(t)$:
\begin{equation}
 \label{dynamic_equation_coupled_oscillator_a_modified}
  \begin{split}
  \frac{dA(t)}{dt} 
& =  \frac{1}{\kappa_{\text{eff}}} \sum_{j} \omega_{j}^{2}  \bigg[ - \omega_{j} \mu_{j} 
  (\theta_{j}(0) - Q(0)) \sin(\omega_{j} t) 
  + \pi_{j}(0) \cos(\omega_{j} t) \\
 & \qquad + \int_{0}^{t} dt' \, 
  \left[ \omega_{j} \mu_{j} \sin(\omega_{j} (t - t')) \frac{dA(t')}{dt'}
   + \cos(\omega_{j}(t - t')) \kappa_{j} (Q(t) - A(t))
   \right] \bigg].
  \end{split}
\end{equation}

We can rewrite eq~\eqref{dynamic_equation_coupled_oscillator_a_modified} in a
simple and intuitive form. 
After some calculations, the dynamic equations become
\begin{align}
 \label{dynamic_equation_coupled_oscillator_transient_potential_q}
 \frac{dQ(t)}{dt} & = \frac{P(t)}{M} , \\
 \label{dynamic_equation_coupled_oscillator_transient_potential_p}
 \frac{dP(t)}{dt} & = - \frac{\partial \Phi(Q(t),A(t))}{\partial Q(t)}, \\
 \label{dynamic_equation_coupled_oscillator_transient_potential_a}
 \frac{dA(t)}{dt} & = - \int_{0}^{t} dt' \, K(t - t') \frac{\partial \Phi(Q(t'),A(t'))}{\partial A(t')}
 + \xi(t) .
\end{align}
(See Appendix~\ref{memory_kernel_and_noise_in_couplsed_oscillator_model} for the detailed calculations.)
$K(t)$ and $\xi(t)$ in eq~\eqref{dynamic_equation_coupled_oscillator_transient_potential_a} can be interpreted as the memory kernel and the noise
for the auxiliary degree of freedom $A(t)$.
We define the Laplace transform of a function $f(t)$ as $f^{*}(s) \equiv \int_{0}^{t} dt \, e^{-t s} f(t)$.
$K(t)$ and $\xi(t)$ are defined via their the Laplace transforms:
\begin{align}
 \label{dynamic_equation_coupled_oscillator_transient_potential_memory_kernel}
 K^{*}(s) & \equiv 
    \left[ 1 + \frac{1}{\kappa_{\text{eff}}} \sum_{j} \frac{\kappa_{j} \omega_{j}^{2} }{\omega_{j}^{2} + s^{2}} \right]^{-1} 
 \frac{1}{\kappa_{\text{eff}}^{2}} \sum_{j} 
  \frac{\kappa_{j} \omega_{j}^{2} s}{\omega_{j}^{2} + s^{2}} , \\
 \label{dynamic_equation_coupled_oscillator_transient_potential_noise}
 \xi^{*}(s) & \equiv 
    \left[ 1 + \frac{1}{\kappa_{\text{eff}}} \sum_{j} \frac{\kappa_{j} \omega_{j}^{2} }{\omega_{j}^{2} + s^{2}} \right]^{-1} 
  \frac{1}{\kappa_{\text{eff}}} \sum_{j} \left[ -   \frac{\kappa_{j} \omega_{j}^{2} }{\omega_{j}^{2} + s^{2}} (\theta_{j}(0) - Q(0)) 
  + \frac{\omega_{j}^{2}  s}{\omega_{j}^{2} + s^{2}} \pi_{j}(0)  \right].
\end{align}
%

As the case in Sec.~\ref{generalized_langevin_equation}, we interpret $\bm{\theta}(0)$ and $\bm{\pi}(0)$ as random
variables and assume that they obey the equilibrium distribution. 
Although we cannot calculate the inverse Laplace transforms of
eqs~\eqref{dynamic_equation_coupled_oscillator_transient_potential_memory_kernel} 
and \eqref{dynamic_equation_coupled_oscillator_transient_potential_noise} analytically,
the first and second order equilibrium moments can be calculated.
We have the following fluctuation-dissipation relation:
\begin{equation}
 \label{fluctuation_dissipation_relation_transient_potential_coupled_oscillator}
  \langle \xi(t) \rangle_{\text{eq},0} = 0, \qquad
 \langle \xi(t) \xi(t') \rangle_{\text{eq},0} = k_{B} T K(|t - t'|) .
\end{equation}
The detailed calculations for eq~\eqref{fluctuation_dissipation_relation_transient_potential_coupled_oscillator}
are shown in Appendix~\ref{memory_kernel_and_noise_in_couplsed_oscillator_model}.

From the calculations shown above, we find that the canonical equations for the coupled oscillator model
(eqs~\eqref{canonical_equation_coupled_oscillator_q_p} and \eqref{canonical_equation_coupled_oscillator_theta_pi}) can be
modified both to the generalized Langevin type equations 
(eqs~\eqref{gle_coupled_oscillator_q} and \eqref{gle_coupled_oscillator_p}),
and to the dynamic equations with the transient potential
(eqs~\eqref{dynamic_equation_coupled_oscillator_transient_potential_q}-\eqref{dynamic_equation_coupled_oscillator_transient_potential_a}).
In both cases, we have employed no approximations and thus
both two sets of dynamic equations (eqs~\eqref{gle_coupled_oscillator_q}
and \eqref{gle_coupled_oscillator_p}, and eqs~\eqref{dynamic_equation_coupled_oscillator_transient_potential_q}-\eqref{dynamic_equation_coupled_oscillator_transient_potential_a})
are exact. We consider that the dynamic
equations with the transient potential can be justified in the same way as
the generalized Langevin type equation.

The results in this section may be interpreted as follows.
The effective dynamic equations depend
on the choice of the coarse-grained degrees of freedom. If we use only $Q(t)$ and $P(t)$ as
the coarse-grained degrees of freedom, we have the generalized Langevin type equations.
If we employ $A(t)$ as the auxiliary degree of
freedom, in addition to $Q(t)$ and $P(t)$, we have the dynamic equations with the
transient potential. We conclude that the generalized Langevin equation is
not the unique expression for the coarse-grained dynamics, and we can
employ the dynamic equations with the transient potential as well.
In Sec.~\ref{theory}, we attempt to derive the dynamic
equations similar to eqs~\eqref{dynamic_equation_coupled_oscillator_transient_potential_q}-\eqref{dynamic_equation_coupled_oscillator_transient_potential_a},
for more general dynamics model.

The dynamic equations with the transient potential
(eqs~\eqref{dynamic_equation_coupled_oscillator_transient_potential_q}-\eqref{dynamic_equation_coupled_oscillator_transient_potential_a})
are not in the same form as the LETP in Ref.~\cite{Uneyama-2020}.
We cannot rewrite eqs~\eqref{dynamic_equation_coupled_oscillator_transient_potential_q}-\eqref{dynamic_equation_coupled_oscillator_transient_potential_a}
into the overdamped Langevin equations.
Thus our result for the coupled oscillator model does not directly justify the LETP.
We consider that some approximations are required to rewrite the dynamic equations
as the LETP. We will discuss how the LETP type dynamic equation can be related to
dynamic equations with the transient potential, later (Sec.~\ref{comparison_with_langevin_equation_with_transient_potential}).

\section{Theory}
\label{theory}

\subsection{Microscopic Dynamics Model}
\label{microscopic_dynamics_model}

We start from a microscopic dynamics model which obeys the Hamiltonian dynamics.
We consider a three dimensional system which consists of $N$ particles.
We express the position and momentum of the $i$-th particle at time $t$ as $\bm{r}_{i}(t)$
and $\bm{p}_{i}(t)$, respectively. Then the state of this system can be fully specified
by two $3 N$-dimensional vectors (or a $6N$-dimensional vector), $\bm{r}(t) \equiv [\bm{r}_{1}(t),\bm{r}_{2}(t),\dots,\bm{r}_{N}(t)]$
and $\bm{p}(t) \equiv [\bm{p}_{1}(t),\bm{p}_{2}(t),\dots,\bm{p}_{N}(t)]$.
The dynamic equations are given as the canonical equations:
\begin{equation}
 \label{canonical_equations_r_p}
 \frac{d\bm{r}(t)}{dt} = \frac{\partial \hat{\mathcal{H}}(\bm{r}(t),\bm{p}(t))}{\partial \bm{p}(t)}, \qquad
 \frac{d\bm{p}(t)}{dt} = - \frac{\partial \hat{\mathcal{H}}(\bm{r}(t),\bm{p}(t))}{\partial \bm{r}(t)},
\end{equation}
where $\hat{\mathcal{H}}(\bm{r},\bm{p})$ is the Hamiltonian.
(In this work, we do not employ a thermostat for the microscopic dynamics.)
For particles interacting via the interaction potential, the Hamiltonian
can be expressed as
\begin{equation}
 \label{hamiltonian_r_p}
  \hat{\mathcal{H}}(\bm{r},\bm{p}) = \frac{1}{2} \bm{p} \cdot \bm{m}^{-1} \cdot \bm{p}
  + \hat{U}(\bm{r}) ,
\end{equation}
with the diagonal mass tensor $\bm{m}$ (which is assumed to be constant) and the interaction potential energy $\hat{U}(\bm{r})$.

In principle, the dynamics of the system can be fully described by
eq~\eqref{canonical_equations_r_p}.
We consider the coarse-graining for this system.
We reduce the degrees of freedom and derive the effective dynamic
equations for coarse-grained variables such as the centers of mass of several particles.
In this work, we limit ourselves to the case where the coarse-grained variables
can be expressed as linear combinations of the original position $\bm{r}(t)$\cite{Uneyama-2020}.
Then we can split the degrees of freedom $\bm{r}(t)$ and $\bm{p}(t)$ into
the coarse-grained slow degrees of freedom and the remaining fast degrees of freedom.
We assume that there is $M$ coarse-grained position variables, and describe
them by an $M$-dimensional vector as $\bm{Q}(t) \equiv [Q_{1}(t),Q_{2}(t),\dots,Q_{M}(t)]$ ($Q_{j}(t)$ is the $j$-th coarse-grained position variable). We describe the momentum for
$\bm{Q}(t)$ as $\bm{P}(t) \equiv [P_{1}(t),P_{2}(t),\dots,P_{M}(t)]$. We express the remaining
$(3 N - M)$ positions by a $(3 N - M)$-dimensional vector as $\bm{\theta}(t) \equiv [\theta_{1}(t),\theta_{2}(t),\dots,\theta_{3 N - M}(t)]$, and
the momentum for $\bm{\theta}(t)$ as $\bm{\pi}(t) \equiv [\pi_{1}(t),\pi_{2}(t),\dots,\pi_{3 N - M}(t)]$.
Then the position in the phase space at time $t$ can be fully specified by
a $6N$-dimensional vector $\bm{\Gamma}(t) \equiv [\bm{Q}(t),\bm{P}(t),\bm{\theta}(t),\bm{\pi}(t)]$.
Although the fast degrees of freedom can be chosen rather arbitrarily,
we choose $\bm{\theta}(t)$ to be
linear combinations of the original positions (in the same way as $\bm{Q}(t)$).
With an appropriate choice of $\bm{\theta}(t)$, the mass tensor can be block-diagonal.
Then we can rewrite the Hamiltonian~\eqref{hamiltonian_r_p} as:
\begin{equation}
 \label{hamiltonian_gamma}
  \hat{\mathcal{H}}(\bm{Q},\bm{P},\bm{\theta},\bm{\pi}) = \frac{1}{2} \bm{P} \cdot \bm{M}^{-1} \cdot \bm{P}
  + \frac{1}{2} \bm{\pi} \cdot \bm{\mu}^{-1} \cdot \bm{\pi}
  + \hat{U}(\bm{Q},\bm{\theta}) ,
\end{equation}
where $\bm{M}$ and $\bm{\mu}$ are mass tensors, and both of them
are assumed to be constant. ($\bm{M}$ and $\bm{\mu}$ are not diagonal, in general.)
The interaction potential $\hat{U}(\bm{Q},\bm{\theta})$ is a function of $\bm{Q}$ and $\bm{\theta}$,
and the fast and slow degrees of freedom are coupled only via the interaction potential.
From eq~\eqref{hamiltonian_gamma}, the canonical equations are explicitly written as
\begin{align}
 \label{canonical_equations_q_p}
 \frac{d\bm{Q}(t)}{dt} 
 & 
 = \bm{M}^{-1} \cdot \bm{P}(t),
 & \frac{d\bm{P}(t)}{dt}
 & = - \frac{\partial \hat{U}(\bm{Q}(t),\bm{\theta}(t))}{\partial \bm{Q}(t)}, \\
 \label{canonical_equations_theta_pi}
 \frac{d\bm{\theta}(t)}{dt} 
& 
 = \bm{\mu}^{-1} \cdot \bm{\pi}(t),
 & \frac{d\bm{\pi}(t)}{dt} 
 & = - \frac{\partial \hat{U}(\bm{Q}(t),\bm{\theta}(t))}{\partial \bm{\theta}(t)}.
\end{align}
Eqs~\eqref{canonical_equations_q_p} and \eqref{canonical_equations_theta_pi} are, of course, equivalent to eq~\eqref{canonical_equations_r_p}.
If we successfully eliminate the degrees of freedom $\bm{\theta}(t)$ and $\bm{\pi}(t)$
from eqs~\eqref{canonical_equations_q_p} and \eqref{canonical_equations_theta_pi},
we will have the coarse-grained effective dynamic equations for $\bm{Q}(t)$ and $\bm{P}(t)$.

We consider a physical quantity which can be expressed as a function of the
position in the $6N$-dimensional phase space.
We express physical quantities
which explicitly depend on $\bm{\theta}(t)$ and/or $\bm{\pi}(t)$ with hats, such as $\hat{f}(\bm{\Gamma}(t))$ and $\hat{g}(\bm{\theta}(t))$.
Some physical quantities depend only on $\bm{Q}(t)$ and $\bm{P}(t)$. We express
physical quantities which depend only on the coarse-grained variables with bars such as
$\bar{h}(\bm{Q}(t),\bm{P}(t))$.

The dynamics of the system can be essentially described by the Poisson
bracket or by the Liouville operator.
The Liouville operator (Liouvillian) $\mathcal{L}$ which
corresponds to canonical equations~\eqref{canonical_equations_q_p} and \eqref{canonical_equations_theta_pi} is given as
\begin{equation}
\begin{split}
 \label{liouville_operator_gamma}
 \mathcal{L} \hat{f}(\bm{\Gamma}(t))
 & \equiv \bm{P}(t) \cdot \bm{M}^{-1} \cdot \frac{\partial \hat{f}(\bm{\Gamma}(t))}{\partial \bm{Q}(t)}
  - \frac{\partial \hat{U}(\bm{Q}(t),\bm{\theta}(t))}{\partial \bm{Q}(t)} \cdot \frac{\partial \hat{f}(\bm{\Gamma}(t))}{\partial \bm{P}(t)} \\
 & \qquad +  \bm{\pi}(t) \cdot \bm{\mu}^{-1} \cdot \frac{\partial \hat{f}(\bm{\Gamma}(t))}{\partial \bm{\theta}(t)}
  - \frac{\partial \hat{U}(\bm{Q}(t),\bm{\theta}(t))}{\partial \bm{\theta}(t)} \cdot \frac{\partial \hat{f}(\bm{\Gamma}(t))}{\partial \bm{\pi}(t)} .
\end{split}
\end{equation}
Since the Hamiltonian dynamics is
deterministic, a time-dependent physical quantity
can be formally expressed as a function of the initial position in the phase space
$\bm{\Gamma}_{0} \equiv \bm{\Gamma}(0)$ and time $t$.
This can be realized by utilizing the time-shift operator. A time-shifted physical quantity can be formally expressed by utilizing the Liouville operator:
$\hat{f}(\bm{\Gamma}(t)) = e^{t \mathcal{L}} \hat{f}(\bm{\Gamma}_{0}) $
(here the operator $\mathcal{L}$ is interpreted as the operator for $\bm{\Gamma}_{0}$).
Then the time derivative of a physical quantity at time $t$ becomes
$(d /  d t) \hat{f}(\bm{\Gamma}(t)) = \mathcal{L} \hat{f}(\bm{\Gamma}(t)) = e^{t \mathcal{L}} \mathcal{L} \hat{f}(\bm{\Gamma}_{0})$.

\subsection{Equilibrium Properties}
\label{equilibrium_properties}

We should eliminate the fast variables $\bm{\theta}(t)$ and $\bm{\pi}(t)$
to obtain the effective dynamic equations for $\bm{Q}(t)$ and $\bm{P}(t)$.
Before we proceed to the elimination of fast variables,
we consider the equilibrium probability distributions.
In this subsection, we consider only static properties. Thus we do
not treat the position in the phase space as a function of time in this subsection, 
and describe variables without the argument $t$, such as $\bm{\Gamma}$ and $\bm{Q}$.
In equilibrium, the canonical probability distribution function in the
$6N$-dimensional phase space can be simply expressed as
\begin{equation}
 \hat{\Psi}_{\text{eq}}(\bm{\Gamma}) = \frac{1}{\mathcal{Z}}
  \exp\left[ - \frac{\hat{\mathcal{H}}(\bm{\Gamma})}{k_{B} T}\right],
\end{equation}
where
$\mathcal{Z}$ is
the partition function defined as follows:
\begin{equation}
 \mathcal{Z} \equiv \int d\bm{\Gamma} \,
  \exp\left[ - \frac{\hat{\mathcal{H}}(\bm{\Gamma})}{k_{B} T}\right].
\end{equation}
The equilibrium probability distribution for the coarse-grained degrees of freedom
can be obtained straightforwardly.
The equilibrium probability distribution for $\bm{Q}$ and $\bm{P}$ simply
becomes
\begin{equation}
 \bar{\Psi}_{\text{eq}}(\bm{Q},\bm{P})
  = \int d\bm{\theta} d\bm{\pi} \, \hat{\Psi}_{\text{eq}}(\bm{\Gamma})
  = \frac{\sqrt{\det (2 \pi \bm{\mu})}}{\mathcal{Z}}
  \exp\left[ - \frac{\bm{P} \cdot \bm{M}^{-1} \cdot \bm{P}}{2 k_{B} T}
       - \frac{\bar{\mathcal{F}}(\bm{Q})}{k_{B} T} \right] ,
\end{equation}
where $\bar{\mathcal{F}}(\bm{Q})$ is the free energy for $\bm{Q}$ defined as
\begin{equation}
 \bar{\mathcal{F}}(\bm{Q}) \equiv
  - k_{B} T \ln \int d\bm{\theta} \, e^{-\hat{U}(\bm{Q},\bm{\theta}) / k_{B} T} .
\end{equation}

In Sec.~\ref{dynamic_equations_with_transient_potential}, we rewrote
the dynamic equations by employing the transient potential and
the additional, auxiliary degree of freedom. Here we introduce the
transient potential to the system in a similar way. We do not know
whether the transient potential for $\bm{Q}$ can be expressed by 
some hypothetical functional forms (such as eq~\eqref{transient_potential_coupled_oscillator}) with auxiliary degrees of freedom or not.
Therefore,
we treat the potential
$\hat{U}(\bm{Q},\bm{\theta})$ itself
as an auxiliary degree of freedom.
We express the transient potential as $\Phi(\tilde{\bm{q}})$
where $\tilde{\bm{q}}$ is a dummy variable.
(In what follows, we employ variables with tildes such as
$\tilde{\bm{q}}$ as dummy variables.)
For static properties, we define the transient potential as
\begin{equation}
 \label{transient_potential_static_definition}
 \Phi(\tilde{\bm{q}}) \equiv \hat{U}(\tilde{\bm{q}},\bm{\theta}),
\end{equation}
and employ $\Phi(\tilde{\bm{q}})$ as an additional degree of freedom.
Here, it should be emphasized that we introduce ``a function'' 
$\Phi(\tilde{\bm{q}})$, not ``a number'' (such as a vector),
as an auxiliary degree of freedom.
Eq~\eqref{transient_potential_static_definition} should hold
for any $\tilde{\bm{q}}$ (and this is why we introduced the dummy variable $\tilde{\bm{q}}$).
Intuitively, eq~\eqref{transient_potential_static_definition} means that
the functional form of $\Phi(\tilde{\bm{q}})$ (not a single value at a specific point such as $\Phi(\bm{Q})$)
should be tuned to reproduce the
original potential form under a given $\bm{\theta}$.
A function $\Phi(\tilde{\bm{q}})$ itself has infinite degrees of freedom,
while $\hat{U}(\tilde{\bm{q}},\bm{\theta})$ has just finite degrees of freedom.
One might think that the introduction of a function with
infinite degrees of freedom may make the expressions complicated and
is useless. However, as we will show, it is useful in some aspects.
This situation would be similar to the introduction of continuum fields
(such as the density field) to particle systems in order to construct approximate
models\cite{Ohta-Kawasaki-1986,Frusawa-Hayakawa-1999,Frusawa-Hayakawa-2000,Matsen-2002,Fredrickson-Ganesan-Drolet-2002,Kawakatsu-book}.

The state of the system at the coarse-grained level can be
now specified by $\bm{Q}, \bm{P}$, and $\Phi(\tilde{\bm{q}})$.
For example, the equilibrium
probability distribution at the coarse-grained level should depend on
$\Phi(\tilde{\bm{q}})$, in addition to
$\bm{Q}$ and $\bm{P}$.
We express physical quantities which depend only on $\bm{Q}$, $\bm{P}$, and $\Phi(\tilde{\bm{q}})$ with bars
such as $\bar{h}[\bm{Q},\bm{P},\Phi(\cdot)]$, in the same way as
the quantities which depend only on $\bm{Q}$ and $\bm{P}$.
Here, $\bar{h}[\bm{Q},\bm{P},\Phi(\cdot)]$ is a function of $\bm{Q}$ and $\bm{P}$
and a functional of $\Phi(\tilde{\bm{q}})$  (``$(\cdot)$'' indicates that the
argument does not represent a specific value of the function at a specific point but the function itself).
The equilibrium distribution function for $\bm{Q}, \bm{P}$, and $\Phi(\tilde{\bm{q}})$
becomes
\begin{equation}
 \label{equilibrium_distribution_q_p_psi}
 \begin{split}
  \bar{\Psi}_{\text{eq}}[\bm{Q},\bm{P},\Phi(\cdot)]
  & = \int d\bm{\theta} d\bm{\pi} \,
  \delta[\Phi(\cdot) - \hat{U}(\cdot,\bm{\theta})] \hat{\Psi}_{\text{eq}}(\bm{\Gamma}) \\
  & = \frac{\sqrt{\det (2 \pi \bm{\mu})}}{\mathcal{Z}}
  \exp\left[ 
  - \frac{1}{k_{B} T} 
  \left[ \frac{1}{2} \bm{P}\cdot\bm{M}^{-1} \cdot \bm{P}
  + \Phi(\bm{Q})
  + \bar{\Upsilon}[\Phi(\cdot)] \right]
      \right] ,
 \end{split}
\end{equation}
where $\delta[\Phi(\cdot) - \hat{U}(\cdot,\bm{\theta})]$ represents the delta functional and $\bar{\Upsilon}[\Phi(\cdot)]$ is defined as
\begin{equation}
 \label{effective_potential_functional_for_transient_potential}
  \bar{\Upsilon}[\Phi(\cdot)] \equiv -k_{B} T \ln  \int d\bm{\theta} \,
  \delta[\Phi(\cdot) - \hat{U}(\cdot,\bm{\theta})] .
\end{equation}
$\bar{\Upsilon}[\Phi(\cdot)]$ would be interpreted as an effective potential
functional for $\Phi(\tilde{\bm{q}})$.
The transient potential $\Phi(\tilde{\bm{q}})$ can take various functional forms, and
the probability to take a specific functional form is determined by eq~\eqref{equilibrium_distribution_q_p_psi}.
If we take the statistical average over all the possible functional forms for
eq~\eqref{equilibrium_distribution_q_p_psi}, we simply have
\begin{equation}
 \label{equilibrium_distribution_q_p_psi_modified}
  \bar{\Psi}_{\text{eq}}(\bm{Q},\bm{P})
  = \int \mathcal{D}\Phi \, \bar{\Psi}_{\text{eq}}[\bm{Q},\bm{P},\Phi(\cdot)].
\end{equation}
Here, $\int \mathcal{D}\Phi$ represents the functional integral over $\Phi(\tilde{\bm{q}})$, and
we have assumed that the measure for the functional integral is appropriately chosen.

The conditional probability distribution (the local equilibrium distribution) of 
the transient potential under given $\bm{Q}$ and $\bm{P}$ becomes
$\bar{\Psi}_{\text{eq}}[\Phi(\cdot) | \bm{Q},\bm{P}]
= \bar{\Psi}_{\text{eq}}[\bm{Q},\bm{P},\Phi(\cdot)] / \bar{\Psi}_{\text{eq}}(\bm{Q},\bm{P})$.
Then we can show that the average force by the transient potential under given $\bm{Q}$ and $\bm{P}$
simply reduces to that by the free energy $\bar{\mathcal{F}}(\bm{Q})$:
\begin{equation}
 \label{equilibrium_average_psi_under_given_q_p}
 \begin{split}
  & \int \mathcal{D}\Phi \,
  \left[ - \frac{\partial \Phi(\bm{Q})}{\partial \bm{Q}} \right]
  \bar{\Psi}_{\text{eq}}[\Phi(\cdot) | \bm{Q},\bm{P}] \\
  & = - \exp\left[  \frac{\bar{\mathcal{F}}(\bm{Q})}{k_{B} T}
      \right]
  \int \mathcal{D} \Phi \int d\bm{\theta} \,
  \delta[\Phi(\cdot) - \hat{U}(\cdot,\bm{\theta})] 
  \frac{\partial \Phi(\bm{Q})}{\partial \bm{Q}}
  \exp\left[  - \frac{\Phi(\bm{Q})}{k_{B} T}
      \right] 
   = - \frac{\partial \bar{\mathcal{F}}(\bm{Q})}{\partial \bm{Q}}.
 \end{split}
\end{equation}
From eqs~\eqref{equilibrium_distribution_q_p_psi} and \eqref{equilibrium_average_psi_under_given_q_p},
in equilibrium, we find that force by the transient potential $- \partial \Phi(\bm{Q}) / \partial \bm{Q}$ fluctuates around 
that by the free energy $- \partial \bar{\mathcal{F}}(\bm{Q}) / \partial \bm{Q}$.

\subsection{Projection Operators and Dynamic Equations for Coarse-Grained Variables}
\label{projection_operators_and_dynamic_equations_for_coarse_grained_variables}

To obtain the effective dynamic equations for $\bm{Q}(t),\bm{P}(t),$ and
the time-dependent transient potential, we 
utilize the projection operator for these variables.
To describe the time-evolution of the transient potential, we
interpret the interaction potential $\hat{U}(\tilde{\bm{q}},\bm{\theta}(t))$ as a function of time (and actually it is a function of time via $\bm{\theta}(t)$).
Then $\hat{U}(\tilde{\bm{q}},\bm{\theta}(t))$ works as the transient potential.
We define the projection operator $\mathcal{P}$ as follows:
\begin{equation}
 \label{projection_operator_p}
 \begin{split}
  \mathcal{P} \hat{f}(\bm{\Gamma}_{0})
  & \equiv \frac{\displaystyle
  \int d\bm{\Gamma}_{0}' \, \hat{\Psi}_{\text{eq}}(\bm{\Gamma}'_{0})
  \delta(\bm{Q}_{0} - \bm{Q}_{0}') \delta(\bm{P}_{0} - \bm{P}_{0}') \delta[\hat{U}(\cdot,\bm{\theta}_{0}) - \hat{U}(\cdot,\bm{\theta}'_{0})]  \hat{f}(\bm{\Gamma}_{0}')}
  {\displaystyle \int d\bm{\Gamma}'_{0} \, \hat{\Psi}_{\text{eq}}(\bm{\Gamma}'_{0})
  \delta(\bm{Q}_{0} - \bm{Q}_{0}') \delta(\bm{P}_{0} - \bm{P}_{0}') \delta[\hat{U}(\cdot,\bm{\theta}_{0}) - \hat{U}(\cdot,\bm{\theta}'_{0})] } \\
  & = \frac{1}
  {\bar{\Psi}_{\text{eq}}[\bm{Q}_{0},\bm{P}_{0},\hat{U}(\cdot,\bm{\theta}_{0})]} 
 \int d\bm{\theta}_{0}' d\bm{\pi}'_{0} \, \hat{\Psi}_{\text{eq}}(\bm{\Gamma}'_{0})
  \delta[\hat{U}(\cdot,\bm{\theta}_{0}) - \hat{U}(\cdot,\bm{\theta}_{0}')] \hat{f}(\bm{\Gamma}_{0}') .
 \end{split}
\end{equation}
Here, 
$\bm{\Gamma}_{0} = [\bm{Q}_{0},\bm{P}_{0},\bm{\theta}_{0},\bm{\pi}_{0}]$
and $\bm{\Gamma}_{0}' = [\bm{Q}_{0}',\bm{P}_{0}',\bm{\theta}_{0}',\bm{\pi}_{0}']$ are
initial positions in the phase space. Eq~\eqref{projection_operator_p}
corresponds to the partial equilibrium average over the initial state,
under fixed $\bm{Q}_{0}, \bm{P}_{0}$,
and $\hat{U}(\tilde{\bm{q}},\bm{\theta}_{0})$.
As before, the functional form of $\hat{U}(\tilde{\bm{q}},\bm{\theta}_{0})$ is
constrained (instead of a specific value such as $\hat{U}(\bm{Q},\bm{\theta}_{0})$)
in eq~\eqref{projection_operator_p}.
Eq~\eqref{projection_operator_p} would be considered as an analog
to the projection operator for the density field in
the time-dependent density functional model\cite{Yoshimori-2005}.
The projection operator $\mathcal{P}$ extracts the component of a physical
quantity
which can be expressed in terms of the initial values of the coarse-grained parameters $\bm{Q}_{0}, \bm{P}_{0}$, and $\hat{U}(\tilde{\bm{q}},\bm{\theta}_{0})$.
From eq~\eqref{projection_operator_p}, it is clear that $\mathcal{P}^{2} = \mathcal{P}$.
We also define another projection operator as $\mathcal{Q} \equiv 1 - \mathcal{P}$
which satisfies $\mathcal{Q}^{2} = \mathcal{Q}$ and $\mathcal{P} \mathcal{Q} = \mathcal{Q} \mathcal{P} = 0$.

The dynamic equations for $\bm{Q}(t)$ and $\bm{P}(t)$ are given by eq~\eqref{canonical_equations_q_p}.
They can be rewritten as
\begin{align}
 \label{time_derivative_q}
 \frac{d}{dt} \bm{Q}(t)
 & 
  = \bm{M}^{-1} \cdot \bm{P}(t) , \\
 \label{time_derivative_p}
 \frac{d}{dt} \bm{P}(t)
 & = e^{t \mathcal{L}} \left[ - \frac{\partial \hat{U}(\bm{Q}_{0},\bm{\theta}_{0})}{\partial \bm{Q}_{0}}\right]
 = - \left. \frac{\partial [e^{t \mathcal{L}} \hat{U}(\tilde{\bm{q}},\bm{\theta}_{0})]}{\partial \tilde{\bm{q}}}
 \right|_{\tilde{\bm{q}} = \bm{Q}(t)}.
\end{align}
Thus we find that, 
if we interpret $e^{t \mathcal{L}} \hat{U}(\tilde{\bm{q}},\bm{\theta}_{0})$
as an additional coarse-grained degree of freedom,
the dynamic
equations for $\bm{Q}(t)$ and $\bm{P}(t)$ can be expressed only with the coarse-grained variables, in rather simple forms.
We can rewrite the dynamic equations in simple forms if we employ the following time-dependent effective Hamiltonian
for coarse-grained variables:
\begin{align}
 \label{time_dependent_effective_hamiltonian_definition}
  \mathcal{H}_{\text{eff}}(\bm{Q},\bm{P},t) & = 
  \frac{1}{2} \bm{P} \cdot \bm{M}^{-1} \cdot \bm{P} + \Phi(\bm{Q},t), \\
 \label{time_dependent_transient_potential_definition}
 \Phi(\tilde{\bm{q}},t)  &  \equiv e^{t \mathcal{L}} \hat{U}(\tilde{\bm{q}},\bm{\theta}_{0}) = \hat{U}(\tilde{\bm{q}},\bm{\theta}(t)).
\end{align}
With the time-dependent effective Hamiltonian \eqref{time_dependent_effective_hamiltonian_definition},
eqs~\eqref{time_derivative_q} and \eqref{time_derivative_p} can be
interpreted as the usual canonical
equations for $\bm{Q}(t)$ and $\bm{P}(t)$.
Eq~\eqref{time_dependent_transient_potential_definition} is the
definition of the time-dependent transient potential. (It reduces to eq~\eqref{transient_potential_static_definition},
if we consider only on the time-independent static properties.)
Intuitively, eq~\eqref{time_dependent_transient_potential_definition} means that
the time-dependent functional form of $\Phi(\tilde{\bm{q}},t)$ is tuned so
that it reproduces the original potential function which evolves in time as $\bm{\theta}(t)$ evolves.
If the time-dependent functional form of
$\Phi(\tilde{\bm{q}},t)$ is given,
the dynamic equations~\eqref{time_derivative_q} and \eqref{time_derivative_p}
are expressed only in terms of the coarse-grained variables
$\bm{Q}(t)$ and $\bm{P}(t)$ at time $t$.
This is much different from the usual case where we 
do not employ the transient potential.
(We will discuss the difference between our method and the standard coarse-graining
method in Sec.~\ref{comparison_with_generalized_langevin_equation}.)

Of course, the dynamic equations \eqref{time_derivative_q} and
\eqref{time_derivative_p} are meaningless unless we explicitly
specify the dynamic equation for the transient potential $\Phi(\tilde{\bm{q}},t)$.
The state of the system at the coarse-grained level
is now specified by $\lbrace \bm{Q}(t),\bm{P}(t),\Phi(\tilde{\bm{q}},t) \rbrace$ instead of $\lbrace \bm{Q}(t),\bm{P}(t) \rbrace$.
We should derive the explicit form of the dynamic equation for $\Phi(\tilde{\bm{q}},t)$.
Following the standard procedure, we use the projection operators $\mathcal{P}$
and $\mathcal{Q}$ to obtain the effective dynamic equation for $\Phi(\tilde{\bm{q}},t)$.
We utilize the following operator identity by Kawasaki\cite{Kawasaki-1973,Dengler-2016}:
\begin{equation}
 \label{kawasaki_operator_identity}
 e^{t \mathcal{L}} = e^{t \mathcal{L}} \mathcal{P}
  + \int_{0}^{t} dt' \, e^{t' \mathcal{L}} \mathcal{P} \mathcal{L}
  \mathcal{Q} e^{(t - t') \mathcal{L} \mathcal{Q}}
  + \mathcal{Q} e^{t \mathcal{L} \mathcal{Q}} .
\end{equation}
By using the operator identity \eqref{kawasaki_operator_identity}
for $(\partial / \partial t) \Phi(\tilde{\bm{q}},t) = e^{t \mathcal{L}} \mathcal{L} \hat{U}(\tilde{\bm{q}},\bm{\theta}_{0})$, we can formally rewrite
the dynamic equation for the transient potential as:
\begin{equation}
 \label{time_derivative_phi_formal}
  \frac{\partial}{\partial t} \Phi(\tilde{\bm{q}},t) 
  = \bar{V}[\bm{Q}(t),\bm{P}(t),\Phi(\cdot,t),\tilde{\bm{q}}]
 + \bar{G}[\bm{Q}(\cdot),\bm{P}(\cdot),\Phi(\cdot,\cdot),\tilde{\bm{q}},t]
 + \hat{\Xi}(\bm{\Gamma}_{0},\tilde{\bm{q}},t) .
\end{equation}
Here, $\bar{V}[\bm{Q}(t),\bm{P}(t),\Phi(\cdot,t),\tilde{\bm{q}}], 
\bar{G}[\bm{Q}(\cdot),\bm{P}(\cdot),\Phi(\cdot,\cdot),\tilde{\bm{q}},t],$ and $\hat{\Xi}(\bm{\Gamma}_{0},\tilde{\bm{q}},t)$ are the reversible (mode-coupling) term,
the damping term, and the fluctuating term, respectively.
The damping term depends on the history of the coarse-grained variables (and thus it is a functional of $\bm{Q}(t)$, $\bm{P}(t)$, and $\Phi(\tilde{\bm{q}},t)$).
They are defined as
\begin{align}
 \label{transient_potential_reversible_term}
 \bar{V}[\bm{Q}(t),\bm{P}(t),\Phi(\cdot,t),\tilde{\bm{q}}]
 & \equiv e^{t \mathcal{L}} \mathcal{P} \mathcal{L} \hat{U}(\tilde{\bm{q}},\bm{\theta}_{0}) ,
 \\
 \label{transient_potential_damping_term}
 \bar{G}[\bm{Q}(\cdot),\bm{P}(\cdot),\Phi(\cdot,\cdot),\tilde{\bm{q}},t]
 & \equiv  \int_{0}^{t} dt' \, e^{t' \mathcal{L}} \mathcal{P} \mathcal{L}
  \mathcal{Q} e^{(t - t') \mathcal{L} \mathcal{Q}}
 \mathcal{L} \hat{U}(\tilde{\bm{q}},\bm{\theta}_{0}),
 \\
 \label{transient_potential_fluctuating_term}
 \hat{\Xi}(\bm{\Gamma}_{0},\tilde{\bm{q}},t) 
 & \equiv  \mathcal{Q} e^{t \mathcal{L} \mathcal{Q}} 
 \mathcal{L} \hat{U}(\tilde{\bm{q}},\bm{\theta}_{0}) .
\end{align}
From eqs~\eqref{transient_potential_reversible_term} and \eqref{transient_potential_damping_term},
the reversible and damping terms can be expressed only in terms of the coarse-grained variables.
On the other hand, from eq~\eqref{transient_potential_fluctuating_term}, clearly
the fluctuating term cannot be expressed only in terms of the coarse-grained variables.

Now we calculate eqs~\eqref{transient_potential_reversible_term}-\eqref{transient_potential_fluctuating_term} separately.
Firstly, we calculate the reversible term $\bar{V}[\bm{Q}(t),\bm{P}(t),\Phi(\cdot,t),\tilde{\bm{q}}]$ (eq~\eqref{transient_potential_reversible_term}):
\begin{equation}
 \label{transient_potential_reversible_term_modified}
 \begin{split}
   \bar{V}[\bm{Q}(t),\bm{P}(t),\Phi(\cdot,t),\tilde{\bm{q}}] 
 & = 
  e^{t \mathcal{L}} \mathcal{P} \left[ \bm{\pi}_{0} \cdot \bm{\mu} \cdot \frac{\partial \hat{U}(\tilde{\bm{q}},\bm{\theta}_{0})}{\partial \bm{\theta}_{0}} \right] \\
  & = e^{t \mathcal{L}} \Bigg[ 
  \frac{1}
  {\bar{\Psi}_{\text{eq}}[\bm{Q}_{0},\bm{P}_{0},\hat{U}(\cdot,\bm{\theta_{0}})]} 
  \int d\bm{\theta}_{0}' \, \delta[\hat{U}(\cdot,\bm{\theta}_{0}') - \hat{U}(\cdot,\bm{\theta}_{0}')] \\
  & \qquad \times \left[  \int d\bm{\pi}_{0}' \, \bm{\pi}_{0}' \hat{\Psi}_{\text{eq}}(\bm{\Gamma}_{0}')  \right] \cdot
  \bm{\mu} \cdot \frac{\partial \hat{U}(\tilde{\bm{q}},\bm{\theta}_{0}')}{\partial \bm{\theta}_{0}'} \Bigg]
  = 0 . 
 \end{split}
\end{equation}
Thus the reversible term is simply zero.
We only have the damping and fluctuating terms for the transient potential.

Secondly, we calculate the damping term $\bar{G}[\bm{Q}(\cdot),\bm{P}(\cdot),\Phi(\cdot,\cdot),\tilde{\bm{q}},t]$ (eq~\eqref{transient_potential_damping_term}).
As Kawasaki explicitly showed\cite{Kawasaki-1973}, the damping term can be expressed in a
simple form by utilizing the memory kernel.
By combining eqs~\eqref{transient_potential_damping_term} and \eqref{transient_potential_fluctuating_term},
the damping term can be rewritten as follows:
\begin{equation}
 \label{transient_potential_damping_term_with_fluctuating_term}
 \bar{G}[\bm{Q}(\cdot),\bm{P}(\cdot),\Phi(\cdot,\cdot),\tilde{\bm{q}},t] 
 =  \int_{0}^{t} dt' \, e^{t' \mathcal{L}} \mathcal{P} \mathcal{L}
  \mathcal{Q} \, \hat{\Xi}(\tilde{\bm{q}},\bm{\Gamma}_{0},t - t').
\end{equation}
Then the damping term reduces to
\begin{equation}
\begin{split}
 \label{transient_potential_damping_term_modified}
 & \bar{G}[\bm{Q}(\cdot),\bm{P}(\cdot),\Phi(\cdot,\cdot),\tilde{\bm{q}},t] \\
 & = k_{B} T \int_{0}^{t} dt' \, e^{t' \mathcal{L}} \Bigg[ \frac{1}{\bar{\Psi}_{\text{eq}}[\bm{Q}_{0},\bm{P}_{0},\hat{U}(\cdot,\bm{\theta}_{0})]}
  \int d\tilde{\bm{q}}' \, \frac{\delta}{\delta \hat{U}(\tilde{\bm{q}}',\bm{\theta}_{0})} \\
 & \qquad \times \left[ \bar{\Psi}_{\text{eq}}
 [\bm{Q}_{0},\bm{P}_{0},\hat{U}(\cdot,\bm{\theta}_{0})]
  \bar{K}[\bm{Q}_{0},\bm{P}_{0},\hat{U}(\cdot,\bm{\theta}_{0}),\tilde{\bm{q}},\tilde{\bm{q}}',t - t']
  \right] \Bigg]  \\
 & = - \int_{0}^{t} dt' \, e^{t' \mathcal{L}} 
  \int d\tilde{\bm{q}}' \, 
  \bar{K}[\bm{Q}_{0},\bm{P}_{0},\hat{U}(\cdot,\bm{\theta}_{0}),\tilde{\bm{q}},\tilde{\bm{q}}',t - t']
\frac{\delta \bar{\Omega}[\bm{Q}_{0},\bm{P}_{0},\hat{U}(\cdot,\bm{\theta}_{0})]}{\delta \hat{U}(\tilde{\bm{q}}',\bm{\theta}_{0})}  \\
 & \qquad + k_{B} T \int_{0}^{t} dt' \, e^{t' \mathcal{L}} 
  \int d\tilde{\bm{q}}' \, \frac{\delta   \bar{K}[\bm{Q}_{0},\bm{P}_{0},\hat{U}(\cdot,\bm{\theta}_{0}),\tilde{\bm{q}},\tilde{\bm{q}}',t - t']}{\delta \hat{U}(\tilde{\bm{q}}',\bm{\theta}_{0})} ,
\end{split}
\end{equation}
where $\delta / \delta \hat{U}(\tilde{\bm{q}},\bm{\theta}_{0})$ represents
the functional differential with respect to $ \hat{U}(\tilde{\bm{q}},\bm{\theta}_{0})$, and
we have defined the memory kernel as
\begin{equation}
 \label{memory_kernel_phi}
 \bar{K}[\bm{Q}_{0},\bm{P}_{0},\hat{U}(\cdot,\bm{\theta}_{0}),\tilde{\bm{q}},\tilde{\bm{q}}',t]
  \equiv \frac{1}{k_{B} T} \mathcal{P} 
  \left[  \hat{\Xi} (\bm{\Gamma}_{0},\tilde{\bm{q}},t)
 \hat{\Xi}(\bm{\Gamma}_{0},\tilde{\bm{q}}',0)\right] ,
\end{equation}
and the free energy functional 
$ \bar{\Omega}[\bm{Q},\bm{P},\Phi(\cdot)]
 \equiv - k_{B} T \ln \bar{\Psi}_{\text{eq}}[\bm{Q},\bm{P},\Phi(\cdot)] $.
(This free energy functional $\bar{\Omega}[\bm{Q},\bm{P},\Phi(\cdot)]$ is different from 
the free energy $\bar{\mathcal{F}}(\bm{Q})$;
$\bar{\Omega}[\bm{Q},\bm{P},\Phi(\cdot)]$ is for  the set of all coarse-grained variables $\lbrace \bm{Q}, \bm{P}, \Phi(\tilde{\bm{q}}) \rbrace$, 
whereas $\bar{\mathcal{F}}(\bm{Q})$ is only for $\bm{Q}$.)
We have utilized several relations for the Liouville and projection operators
to calculate eq~\eqref{transient_potential_damping_term_modified}.
The detailed calculations are shown in Appendix~\ref{some_relations_which_involve_liouville_and_projection_operators}.
From eq~\eqref{equilibrium_distribution_q_p_psi}, the free energy functional $\bar{\Omega}[\bm{Q},\bm{P},\Phi(\cdot)]$ can be rewritten as the following simple form:
\begin{equation}
 \label{free_energy_q_p_psi}
  \bar{\Omega}[\bm{Q},\bm{P},\Phi(\cdot)]
   = \frac{1}{2} \bm{P}\cdot\bm{M}^{-1} \cdot \bm{P} + \Phi(\bm{Q})
  + \bar{\Upsilon}[\Phi(\cdot)] + (\text{const.}).
\end{equation}

Thirdly, we consider the property of the fluctuating term
$\hat{\Xi}(\bm{\Gamma}_{0},\tilde{\bm{q}},t)$ (eq~\eqref{transient_potential_fluctuating_term}). The fluctuating term can be
interpreted as a random noise, from the viewpoint of the coarse-grained variables.
(From the microscopic viewpoint, $\hat{\Xi}(\bm{\Gamma}_{0},\tilde{\bm{q}},t)$ is 
fully determined by the initial state $\bm{\Gamma}_{0}$ and is not a noise.)
From eq~\eqref{transient_potential_fluctuating_term}, clearly the fluctuating term 
is perpendicular to the coarse-grained variables:
$ \mathcal{P} \hat{\Xi}(\bm{\Gamma}_{0},\tilde{\bm{q}},t) = 0$.
This is consistent with our intuitive picture of the noise.
To characterize the properties of the fluctuating term, the (full) equilibrium statistical average
is suitable. We define the equilibrium statistical average of a physical quantity $\hat{f}(\bm{\Gamma}_{0})$
over the initial state as:
\begin{equation}
 \left\langle \hat{f}(\bm{\Gamma}_{0}) \right\rangle_{\text{eq},0}
  \equiv \int d\bm{\Gamma}_{0} \, \hat{\Psi}_{\text{eq}}(\bm{\Gamma}_{0})  \hat{f}(\bm{\Gamma}_{0}).
\end{equation}
The equilibrium statistical average of $\hat{\Xi}(\bm{\Gamma}_{0},\tilde{\bm{q}},t)$ is zero:
\begin{equation}
 \label{transient_potential_fluctuating_term_average_first_order_moment}
 \left\langle \hat{\Xi}(\bm{\Gamma}_{0},\tilde{\bm{q}},t)
  \right\rangle_{\text{eq},0}
  = \int d\bm{\Gamma}_{0} \, \hat{\Psi}_{\text{eq}}(\bm{\Gamma}_{0})
  \mathcal{P} \hat{\Xi}(\bm{\Gamma}_{0},\tilde{\bm{q}},t)= 0.
\end{equation}
The equilibrium statistical average of the second order moment can be
calculated as
\begin{equation}
 \label{transient_potential_fluctuating_term_average_second_order_moment_symmetrized}
   \left\langle\hat{\Xi}(\bm{\Gamma}_{0},\tilde{\bm{q}},t)
   \hat{\Xi}(\bm{\Gamma}_{0},\tilde{\bm{q}}',t')
  \right\rangle_{\text{eq},0} 
   = k_{B} T
  \left\langle 
  \bar{K}[\bm{Q}_{0},\bm{P}_{0},\hat{U}(\cdot,\bm{\theta}_{0}),\tilde{\bm{q}},\tilde{\bm{q}}',|t - t'|] \right\rangle_{\text{eq},0} ,
\end{equation}
(The detailed calculations are shown in Appendix~\ref{some_relations_which_involve_liouville_and_projection_operators}.)
From eq~\eqref{transient_potential_fluctuating_term_average_second_order_moment_symmetrized}, the average
of the second order moment can be related to the memory kernel \eqref{memory_kernel_phi}.
Eqs~\eqref{transient_potential_fluctuating_term_average_first_order_moment}
and \eqref{transient_potential_fluctuating_term_average_second_order_moment_symmetrized}
can be interpreted as the fluctuation-dissipation relation of the second kind.

From the results shown above,
we can describe the dynamic equation for $\Phi(\tilde{\bm{q}},t)$ in an explicit form.
The resulting dynamic equation is not a closed form in $\lbrace \bm{Q}(t), \bm{P}(t), \Phi(\tilde{\bm{q}},t) \rbrace$, because the fluctuating
term $\hat{\Xi}(\bm{\Gamma}_{0},\tilde{\bm{q}},t)$ cannot be expressed in terms of
coarse-grained variables. Therefore,
we interpret $\hat{\Xi}(\bm{\Gamma}_{0},\tilde{\bm{q}},t)$
as a random noise field, and rewrite the dynamic equation for $\Phi(\tilde{\bm{q}},t)$
as a stochastic dynamic equation\cite{vanKampen-book}. The time origin $t = 0$ can be took arbitrarily and
thus we change the time integral from $0$ to $t$ to that from $-\infty$ to $t$.
We rewrite the noise field as $\Xi[\bm{Q}(\cdot),\bm{P}(\cdot),\Phi(\cdot,\cdot),\tilde{\bm{q}},t]$,
and assume that it is a stochastic process.
In general, the noise field $\Xi[\bm{Q}(\cdot),\bm{P}(\cdot),\Phi(\cdot,\cdot),\tilde{\bm{q}},t]$ depends on
the history of the coarse-grained variables. Thus we should interpret the noise field as the multiplicative colored noise.
By combining eqs~\eqref{time_derivative_q}, \eqref{time_derivative_p}, \eqref{time_derivative_phi_formal},
\eqref{transient_potential_reversible_term_modified},
and \eqref{transient_potential_damping_term_modified},
finally we have the effective dynamic equations
for the coarse-grained variables $\lbrace \bm{Q}(t),\bm{P}(t),\Phi(\tilde{\bm{q}},t)\rbrace$:
\begin{align}
 \label{effective_dynamic_equation_q}
  \frac{d\bm{Q}(t)}{dt}
 & = \bm{M}^{-1} \cdot \bm{P}(t) , \\
 \frac{d\bm{P}(t)}{dt} 
 \label{effective_dynamic_equation_p}
 &  = - \frac{\partial \Phi(\bm{Q}(t),t)}{\partial \bm{Q}(t)} , \\
 \label{effective_dynamic_equation_phi}
\begin{split}
   \frac{\partial \Phi(\tilde{\bm{q}},t)}{\partial t} 
 & = - \int_{-\infty}^{t} dt'
 d\tilde{\bm{q}}' \, 
  \bar{K}[\bm{Q}(t'),\bm{P}(t'),\Phi(\cdot,t'),\tilde{\bm{q}},\tilde{\bm{q}}',t - t']
 \frac{\delta \bar{\Omega}[\bm{Q}(t'),\bm{P}(t'),\Phi(\cdot,t')]}{\delta \Phi(\tilde{\bm{q}}',t')}  \\
 & \qquad + k_{B} T \int_{-\infty}^{t} dt' d\tilde{\bm{q}}' \, 
 \frac{\delta   \bar{K}[\bm{Q}(t'),\bm{P}(t'),\Phi(\cdot,t'),\tilde{\bm{q}},\tilde{\bm{q}}',t - t']}{\delta \Phi(\tilde{\bm{q}}',t')} \\
 & \qquad + \Xi[\bm{Q}(\cdot),\bm{P}(\cdot),\Phi(\cdot,\cdot),\tilde{\bm{q}},t].
\end{split}
\end{align} 
We interpret $\langle \dots \rangle_{\text{eq},0}$ as the equilibrium statistical average
over the coarse-grained variables $\lbrace \bm{Q},\bm{P},\Phi(\tilde{\bm{q}}) \rbrace$
(with the distribution function \eqref{equilibrium_distribution_q_p_psi})
and also over the stochastic process.
Then the first and second moments of $\Xi(\tilde{\bm{q}},t)$ 
(eqs~\eqref{transient_potential_fluctuating_term_average_first_order_moment} and
\eqref{transient_potential_fluctuating_term_average_second_order_moment_symmetrized}) are rewritten as
\begin{equation}
 \label{effective_fluctuation_dissipation_relation_xi_1st}
 \left\langle \Xi[\bm{Q}(\cdot),\bm{P}(\cdot),\Phi(\cdot,\cdot),\tilde{\bm{q}},t] \right\rangle_{\text{eq}} 
  = 0, 
\end{equation}
\begin{equation}
 \label{effective_fluctuation_dissipation_relation_xi_2nd}
  \begin{split}
   & \left\langle \Xi[\bm{Q}(\cdot),\bm{P}(\cdot),\Phi(\cdot,\cdot),\tilde{\bm{q}},t]
   \, \Xi[\bm{Q}(\cdot),\bm{P}(\cdot),\Phi(\cdot,\cdot),\tilde{\bm{q}}',t'] \right\rangle_{\text{eq}} \\
 & = k_{B} T \left\langle \bar{K}[\bm{Q},\bm{P},\Phi(\cdot),\tilde{\bm{q}},\tilde{\bm{q}}',|t - t'|] \right\rangle_{\text{eq}} .
  \end{split} 
\end{equation}
Here, $\langle \dots \rangle_{\text{eq}}$ represents the equilibrium statistical average
at the coarse-grained level.
It should be noticed that only the first and second order moments
for the noise field $\Xi[\bm{Q}(\cdot),\bm{P}(\cdot),\Phi(\cdot,\cdot),\tilde{\bm{q}},t]$
are given.
The noise field $\Xi[\bm{Q}(\cdot),\bm{P}(\cdot),\Phi(\cdot,\cdot),\tilde{\bm{q}},t]$ is not necessarily to be
Gaussian.

Eqs~\eqref{effective_dynamic_equation_q}-\eqref{effective_fluctuation_dissipation_relation_xi_2nd}
are one of the main results of this work.
We conclude that we can formally construct the dynamic equations
with transient potential from the microscopic Hamiltonian dynamics, without any approximations.
The dynamic equation for the transient potential (eq~\eqref{effective_dynamic_equation_phi})
can be interpreted as the generalized Langevin equation for the transient potential.
Therefore, at least formally, it inherits the thermodynamic properties of the generalized
Langevin equation.
It would be
fair to mention that eq~\eqref{effective_dynamic_equation_phi} is just a formal equation.
It is a stochastic partial differential equation in the $M$-dimensional space with the memory effect,
and is useless for practical purposes.
In some aspects, Eq~\eqref{effective_dynamic_equation_phi} would be similar to the stochastic
dynamic density functional theory\cite{Dean-1996,Frusawa-Hayakawa-2000,Archer-Rauscher-2004} where the dynamics
of the system is described by a stochastic partial differential equation for the density field.
The fact that eqs~\eqref{effective_dynamic_equation_q}-\eqref{effective_fluctuation_dissipation_relation_xi_2nd} are just formal does not mean that
they are meaningless.
They justify the concept of the transient potential from the viewpoint of the microscopic Hamiltonian dynamics.
They can be also utilized as a starting point to construct approximate yet simpler
dynamic equations. In Secs.~\ref{markov_approximation} and \ref{approximation_by_potential_parameters},
we introduce some approximations and simplify the dynamic equations.
%

\subsection{Markov Approximation}
\label{markov_approximation}

In Sec.~\ref{projection_operators_and_dynamic_equations_for_coarse_grained_variables},
we have shown that the transient potential obeys the
stochastic partial differential equation with the memory kernel (eq~\eqref{effective_dynamic_equation_phi}).
Such a partial stochastic differential
equation with the memory kernel is clearly difficult to handle with.
We want to approximate eq~\eqref{effective_dynamic_equation_phi} by some
simpler forms. One possible approximation is to ignore the memory effect
(the Markov approximation).
The memory kernel in eq~\eqref{effective_dynamic_equation_phi}
can be approximately eliminated,
if the characteristic time scale
of the memory is sufficiently short
compared with that of the coarse-grained variables.
Naively, we expect that the memory kernel reflects the dynamics
of the eliminated fast degrees of freedom.
Then, as far as the fast degrees of freedom move much faster than
the coarse-grained degrees of freedom, the time scales can be assumed
to be well separated and the characteristic time scale of the memory
is short.
Although whether this assumption is really reasonable or not is not fully clear,
here we simply employ it.

Therefore we approximate the memory kernel in eq~\eqref{effective_dynamic_equation_phi} as
\begin{equation}
 \bar{K}[\bm{Q},\bm{P},\Phi(\cdot),\tilde{\bm{q}},\tilde{\bm{q}}',t] 
  \approx 
  2 \bar{L}[\bm{Q},\bm{P},\Phi(\cdot),\tilde{\bm{q}},\tilde{\bm{q}}'] \delta(t),
\end{equation}
with the memory-less mobility defined as

\begin{equation}
 \label{mobility_phi_markovian}
  \bar{L}[\bm{Q},\bm{P},\Phi(\cdot),\tilde{\bm{q}},\tilde{\bm{q}}']
  \equiv \int_{0}^{\infty} dt \, \bar{K}[\bm{Q},\bm{P},\Phi(\cdot),\tilde{\bm{q}},\tilde{\bm{q}}',t] .
\end{equation}
Then we have the following stochastic partial differential equation without the memory effect
as an approximation for eq~\eqref{effective_dynamic_equation_phi}:
\begin{equation}
 \label{effective_dynamic_equation_phi_markovian}
\begin{split}
 \frac{\partial \Phi(\tilde{\bm{q}},t)}{\partial t} 
 & \approx - 
 \int d\tilde{\bm{q}}' \, \bar{L}[\bm{Q}(t),\bm{P}(t),\Phi(\cdot,t),\tilde{\bm{q}},\tilde{\bm{q}}']
 \frac{\delta \bar{\Omega}[\bm{Q}(t),\bm{P}(t),\Phi(\cdot,t)]}{\delta \Phi(\tilde{\bm{q}}',t)}  \\
 & \qquad + k_{B} T \int d\tilde{\bm{q}}' \, 
 \frac{\delta   \bar{L}[\bm{Q}(t),\bm{P}(t),\Phi(\cdot,t),\tilde{\bm{q}},\tilde{\bm{q}}']}{\delta \Phi(\tilde{\bm{q}}',t)}  \\
 & \qquad + \sqrt{2 k_{B} T}
  \int d\tilde{\bm{q}}' \, \bar{B}[\bm{Q}(t),\bm{P}(t),\Phi(\cdot,t),\tilde{\bm{q}},\tilde{\bm{q}}']
 W(\tilde{\bm{q}}',t).
\end{split}
\end{equation}
To derive eq~\eqref{effective_dynamic_equation_phi_markovian},
we have assumed that $\int_{0}^{\infty} dt \, \delta(t) = 1/2$.
$\bar{B}[\bm{Q},\bm{P},\Phi(\cdot),\tilde{\bm{q}},\tilde{\bm{q}}']$ is the noise coefficient and
$W(\tilde{\bm{q}},t)$ is the white noise field.
The noise coefficient $\bar{B}[\bm{Q},\bm{P},\Phi(\cdot),\tilde{\bm{q}},\tilde{\bm{q}}']$ is defined via
the following relation:
\begin{equation}
  \bar{L}[\bm{Q},\bm{P},\Phi(\cdot),\tilde{\bm{q}},\tilde{\bm{q}}']
  \equiv \int d\tilde{\bm{q}}''
 \bar{B}[\bm{Q},\bm{P},\Phi(\cdot),\tilde{\bm{q}},\tilde{\bm{q}}'']
 \bar{B}[\bm{Q},\bm{P},\Phi(\cdot),\tilde{\bm{q}}',\tilde{\bm{q}}''] ,
\end{equation}
The first and second order moments of $W(\tilde{\bm{q}},t)$ are given as
\begin{equation}
 \label{effective_fluctuation_dissipation_relation_xi_markovian}
 \left\langle W(\tilde{\bm{q}},t) \right\rangle_{\text{eq}} = 0, \qquad
 \left\langle W(\tilde{\bm{q}},t)  W(\tilde{\bm{q}}',t') \right\rangle_{\text{eq}}
 = \delta(t - t') \delta(\tilde{\bm{q}} - \tilde{\bm{q}}').
\end{equation}

Eqs~\eqref{effective_dynamic_equation_q},
\eqref{effective_dynamic_equation_p}, and \eqref{effective_dynamic_equation_phi_markovian}
are the approximate Markovian dynamic equations with the transient potential.
Here, we should recall that the noise field $W(\tilde{\bm{q}},t)$ is not necessarily to be Gaussian.
Therefore, the approximate dynamic equation 
\eqref{effective_dynamic_equation_phi_markovian} is Markovian but not
Gaussian, in general.
For some practical purposes, we may assume the noise field to be Gaussian (the Gaussian noise
approximation).
As we mentioned, eq~\eqref{effective_dynamic_equation_phi} can be
interpreted as the generalized Langevin equation for the transient potential.
Then, eq~\eqref{effective_dynamic_equation_phi_markovian} can be
interpreted as the (overdamped) Langevin equation for the transient potential.
It inherits the thermodynamic properties of the (usual) overdamped Langevin equation,
such as the symmetric and positive definite nature of the mobility $\bar{L}[\bm{Q},\bm{P},\Phi(\cdot),\tilde{\bm{q}},\tilde{\bm{q}}']$
and the detailed-balance condition.

The Markovian dynamic equation \eqref{effective_dynamic_equation_phi_markovian}
is much easier to handle with, compared with the non-Markovian dynamic equation
\eqref{effective_dynamic_equation_phi} with the memory kernel. It should be emphasized
here that we still have the memory effect for the coarse-grained variables $\bm{Q}(t)$ and $\bm{P}(t)$
even under the Markov approximation for $\Phi(\tilde{\bm{q}},t)$. This is because
the dynamic equation for $\bm{P}(t)$ depends on $\Phi(\tilde{\bm{q}},t)$.
$\Phi(\tilde{\bm{q}},t)$ obeys its own dynamic equation \eqref{effective_dynamic_equation_phi_markovian}
and thus it has the time correlation. This time correlation will be observed
as the memory effect for $\bm{P}(t)$.
Therefore, if we observe only $\bm{Q}(t)$ and $\bm{P}(t)$, we
will find that the dynamics is non-Markovian.
The situation would be somewhat similar to the hidden Markov model\cite{Rabiner-1989}.
Such an expression of the memory effect is one of the original
motivation to introduce the transient potential
in the phenomenological coarse-grained dynamics models\cite{Kindt-Briels-2007,Briels-2009,Padding-Briels-2011}.
We consider that our dynamic equations are able to reproduce complex dynamics
even under the Markov approximation, as expected, and justify the phenomenological
use of the transient potential to reproduce the memory effect.

\subsection{Approximation by Potential Parameters}
\label{approximation_by_potential_parameters}

In Sec.~\ref{markov_approximation}, The dynamic equation for $\Phi(\tilde{\bm{q}},t)$ was simplified as
eq~\eqref{effective_dynamic_equation_phi_markovian}
under the Markov approximation. Unfortunately,
even under the Markov approximation, the dynamic equation is still complicated. The analytic handling
as well as numerical simulations are still difficult.
The reason is that eq~\eqref{effective_dynamic_equation_phi_markovian}
is a stochastic partial differential equation for a function in the $M$-dimensional
space.
A function which has infinite degrees of freedom.
This makes theoretical analyses difficult.
Even if we discretize the transient potential,
it still has larger degrees of freedom than the microscopic model.
Thus the numerical simulations would be practically impossible.
Even if we can perform simulations, it would be highly memory-consuming and
inefficient.
Therefore, we want to further simplify the dynamic equation for
$\Phi(\tilde{\bm{q}},t)$ with some additional approximations.

In Sec.~\ref{dynamic_equations_with_transient_potential}, the dynamics
of the transient potential was not expressed as the dynamic equation for the transient potential itself,
but as the dynamic equation for the effective position. If we can
approximately express the transient potential by finite parameters, the
dynamics of the transient potential will be described in a much simpler form.
Such an approximation was also proposed in the previous work\cite{Uneyama-2020}.
Here we consider a
possible approximation for
the transient potential by finite parameters.
We assume a hypothetical form for the transient potential, and try to mimic the full dynamics of the transient potential.
We approximate the transient potential by
a specific hypothetical functional form with time-dependent potential parameters $\bm{A}(t)$: $\Phi(\tilde{\bm{q}},t) \approx \check{\Phi}(\tilde{\bm{q}},\bm{A}(t))$.
This procedure would be interpreted as the variational method. The hypothetical
transient potential $\check{\Phi}(\tilde{\bm{q}},\bm{A}(t))$ is a trial function,
and the parameter $\bm{A}(t)$ is determined so that the error between the true
and trial transient potentials is minimized.
(We will discuss a possible method to estimate the potential parameters from
a given transient potential and a hypothetical transient potential form, in
Sec.~\ref{estimating_potential_parameters_from_transient_potential}.)

Once the hypothetical functional form is given,
the degrees of freedom of the system is changed from $\lbrace \bm{Q}(t),\bm{P}(t),\Phi(\tilde{\bm{q}},t) \rbrace$
to $\lbrace \bm{Q}(t),\bm{P}(t),\bm{A}(t) \rbrace$.
As we explained, the new auxiliary degrees of freedom $\bm{A}(t)$ represent potential parameters in the
hypothetical transient potential function, and
they behave as usual coarse-grained degrees of freedom (such as $\bm{Q}(t)$ and $\bm{P}(t)$). The dimension of
$\bm{A}(t)$ can be arbitrarily chosen (depending on the functional form of $\check{\Phi}(\tilde{\bm{q}},\bm{A}(t))$).
 Here we assume that $\bm{A}(t)$ is a $Z$-dimensional vector: $\bm{A}(t) = [A_{1}(t),A_{2}(t),\dots,A_{Z}(t)]$.
In what follows, the physical quantities which depend on the potential parameters $\bm{A}(t)$
are expressed with checks such as $\check{f}(\bm{A}(t))$.
(In the previous work\cite{Uneyama-2020}, $\bm{A}(t)$ was referred as the pseudo thermodynamic
degrees of freedom because they behave just in the same way as the usual
coarse-grained degrees of freedom which determine the thermodynamic state.)

If we employ the hypothetical transient potential $\check{\Phi}(\tilde{\bm{q}},\bm{A}(t))$,
the equilibrium probability distribution should be expressed as a function 
of $\bm{Q}$, $\bm{P}$, and $\bm{A}$ as $\check{\Psi}_{\text{eq}}(\bm{Q},\bm{P},\bm{A})$.
We assume that the hypothetical transient potential
works as a reasonable approximation for
the original transient potential. Then, we expect that some relations which involve $\Phi(\tilde{\bm{q}})$
should be replaced by similar relations which involve $\bm{A}$.
In analogy to eq~\eqref{equilibrium_distribution_q_p_psi_modified},
$\check{\Psi}_{\text{eq}}(\bm{Q},\bm{P},\bm{A})$ should be related to 
$\bar{\Psi}_{\text{eq}}(\bm{Q},\bm{P})$
via the following relation:
\begin{equation}
 \label{equilibrium_distribution_q_p_a}
 \Psi_{\text{eq}}(\bm{Q},\bm{P})
  = \int d\bm{A} \, \check{\Psi}_{\text{eq}}(\bm{Q},\bm{P},\bm{A})
\end{equation}
Also, the statistical average of the force by the transient potential should reduce to
that by the free energy in a similar way to eq~\eqref{equilibrium_average_psi_under_given_q_p}:
\begin{equation}
 \label{equilibrium_average_psi_hypothetical_under_given_q_p}
  - \frac{\partial \bar{\mathcal{F}}(\bm{Q})}{\partial \bm{Q}} = \int d\bm{A} \,
  \left[ - \frac{\partial\check{\Phi}(\bm{Q},\bm{A})}{\partial \bm{Q}} \right]
   \check{\Psi}_{\text{eq}}(\bm{A}|\bm{Q},\bm{P}),
\end{equation}
with $\check{\Psi}_{\text{eq}}(\bm{A}|\bm{Q},\bm{P})
\equiv \check{\Psi}_{\text{eq}}(\bm{Q},\bm{P},\bm{A}) / \bar{\Psi}_{\text{eq}}(\bm{Q},\bm{P})$.
The functional form of $\check{\Phi}(\tilde{\bm{q}},\bm{A})$ should be selected so that
eqs~\eqref{equilibrium_distribution_q_p_a} and \eqref{equilibrium_average_psi_hypothetical_under_given_q_p}
hold (at least approximately). The free energy
for $\bm{Q}$, $\bm{P}$, and $\bm{A}$ would be expressed as
\begin{equation}
 \label{free_energy_q_p_a}
\begin{split}
   \check{\Omega}(\bm{Q},\bm{P},\bm{A})
 & \equiv - k_{B} T \ln \check{\Psi}_{\text{eq}}(\bm{Q},\bm{P},\bm{A}) \\
 & = \frac{1}{2} \bm{P} \cdot \bm{M}^{-1} \cdot \bm{P}
  + \check{\Phi}(\bm{Q},\bm{A}) + \check{\Upsilon}(\bm{A}) + (\text{const.}),
\end{split}
\end{equation}
where we have introduced the effective potential $\check{\Upsilon}(\bm{A})$ which depends
only on $\bm{A}$.
Here it should be noticed that $\check{\Upsilon}(\bm{A})$ does not corresponds
to the approximation for ${\Upsilon}[\Phi(\cdot)]$ defined by eq~\eqref{effective_potential_functional_for_transient_potential}. This is because the integral is taken over
the function $\Phi(\tilde{\bm{q}})$ in eq~\eqref{equilibrium_distribution_q_p_psi_modified}
whereas it is taken over the vector $\bm{A}$
in eq~\eqref{equilibrium_distribution_q_p_a}.
The variable transform gives additional factor to the free energy arising from
the Jacobian\cite{Nakamura-2018,Uneyama-2020a}.

By assuming the relation $\Phi(\tilde{\bm{q}},t) \approx \check{\Phi}(\tilde{\bm{q}},\bm{A}(t))$, we can approximately rewrite the
derivatives as follows:
\begin{align}
 \label{time_derivative_phi_to_a}
 \frac{\partial \Phi(\tilde{\bm{q}},t)}{\partial t}
 & \approx 
  \check{\bm{J}}(\tilde{\bm{q}},\bm{A}(t)) \cdot \frac{d\bm{A}(t)}{dt} , \\
 \label{partial_derivative_phi_to_a}
 \frac{\partial}{\partial \bm{A}(t)}
 & \approx \int d\tilde{\bm{q}} \, \check{\bm{J}}(\tilde{\bm{q}},\bm{A}(t)) \frac{\delta}{\delta \Phi(\tilde{\bm{q}},t)} ,
\end{align}
where $\check{\bm{J}}(\tilde{\bm{q}},\bm{A}) \equiv \partial \check{\Phi}(\tilde{\bm{q}},\bm{A}) / \partial \bm{A}$.
Eqs~\eqref{time_derivative_phi_to_a} and \eqref{partial_derivative_phi_to_a}
cannot be inverted since $\check{\bm{J}}(\tilde{\bm{q}},\bm{A})$ cannot
be inverted in general. Even though the exact inversion is not possible, we can still
seek an approximate inversion. We employ the Moore-Penrose pseudo inverse\cite{BenIsrael-Greville-book}
to approximately invert eqs~\eqref{time_derivative_phi_to_a} and
\eqref{partial_derivative_phi_to_a}\cite{Uneyama-2020a}. The Moore-Penrose pseudo inverse 
$\check{\bm{J}}^{+}(\tilde{\bm{q}},\bm{A})$ is defined via
$\int d\tilde{\bm{q}}' \, \check{\bm{J}}(\tilde{\bm{q}},\bm{A}) \cdot \check{\bm{J}}^{+}(\tilde{\bm{q}}',\bm{A})
\check{\bm{J}}(\tilde{\bm{q}}',\bm{A}) = \check{\bm{J}}(\tilde{\bm{q}},\bm{A}) $.
The Moore-Penrose pseudo inverse gives the most reasonable solution
for an inversion problem without a unique solution. Thus, we may employ
it as a physically reasonable approximate inverse
and invert eqs~\eqref{time_derivative_phi_to_a} and \eqref{partial_derivative_phi_to_a} as:
\begin{align}
 \frac{d\bm{A}(t)}{dt} 
 & \approx
 \int d\tilde{\bm{q}} \, \check{\bm{J}}^{+}(\tilde{\bm{q}},\bm{A}(t)) \frac{\partial \Phi(\tilde{\bm{q}},t)}{\partial t} , \\
 \frac{\delta}{\delta \Phi(\tilde{\bm{q}},t)}
 & \approx  \check{\bm{J}}^{+}(\tilde{\bm{q}},\bm{A}(t)) \cdot \frac{\partial}{\partial \bm{A}(t)} .
\end{align}
Then the dynamic equation for the transient potential $\Phi(\tilde{\bm{q}},t)$ 
(eq~\eqref{effective_dynamic_equation_phi_markovian}) can be
approximately converted to the dynamic equation for $\bm{A}(t)$:
\begin{equation}
 \label{approximate_effective_dynamic_equation_a_final}
\begin{split}
 \frac{d \bm{A}(t)}{d t} 
 & \approx - 
  \check{\bm{L}}(\bm{Q}(t),\bm{P}(t),\bm{A}(t)) \cdot
 \left[ 
 \frac{\partial \check{\Phi}(\bm{Q}(t),\bm{A}(t))}{\partial \bm{A}(t)} 
 + \frac{\partial \check{\Upsilon}(\bm{A}(t))}{\partial \bm{A}(t)} 
 \right] \\
 & \qquad + k_{B} T \frac{\partial}{\partial \bm{A}(t)} \cdot 
 \check{\bm{L}}(\bm{Q}(t),\bm{P}(t),\bm{A}(t)) + \sqrt{2 k_{B} T}
 \check{\bm{B}}(\bm{Q}(t),\bm{P}(t),\bm{A}(t)) \cdot
 \bm{W}(t),
\end{split}
\end{equation}
where $\check{\bm{L}}(\bm{Q},\bm{P},\bm{A})$ and $\check{\bm{B}}(\bm{Q},\bm{P},\bm{A})$ are
the mobility and the noise coefficient tensors for $\bm{A}(t)$,
and $\bm{W}(t)$ is the $Z$-dimensional white noise vector.
Both $\check{\bm{L}}(\bm{Q},\bm{P},\bm{A})$ and $\check{\bm{B}}(\bm{Q},\bm{P},\bm{A})$
are 
second order tensors with $Z \times Z$ elements and are defined as
\begin{equation}
 \label{approximate_mobility_a}
\begin{split}
  \check{\bm{L}}(\bm{Q},\bm{P},\bm{A}) 
  & \equiv
 \int d\tilde{\bm{q}} d\tilde{\bm{q}}' \,
   \check{\bm{J}}^{+}(\tilde{\bm{q}},\bm{A})  \check{\bm{J}}^{+}(\tilde{\bm{q}}',\bm{A})
  \bar{L}[\bm{Q},\bm{P},\check{\Phi}(\cdot,\bm{A}),\tilde{\bm{q}},\tilde{\bm{q}}'] \\
 & \equiv \check{\bm{B}}(\bm{Q},\bm{P},\bm{A}) \cdot
   \check{\bm{B}}^{\mathrm{T}}(\bm{Q},\bm{P},\bm{A}) ,
\end{split}
\end{equation}
where $\check{\bm{B}}^{\mathrm{T}}(\bm{Q},\bm{P},\bm{A})$ represents the transpose of $\check{\bm{B}}(\bm{Q},\bm{P},\bm{A})$.
The first and second order moments of the noise vector
$\bm{W}(t)$ are
\begin{equation}
 \label{fluctuation_dissipation_relation_approximate}
 \langle \bm{W}(t) \rangle_{\text{eq}} = 0, \qquad
  \langle \bm{W}(t) \bm{W}(t') \rangle_{\text{eq}} = \delta(t - t') \bm{1},
\end{equation}
where $\bm{1}$ is the unit tensor.
The noise vector $\bm{W}(t)$ is not necessarily to be Gaussian, in general.
We expect that the approximate dynamic equation~\eqref{approximate_effective_dynamic_equation_a_final}
inherits the thermodynamic properties of eq~\eqref{effective_dynamic_equation_phi_markovian}; the mobility by eq~\eqref{approximate_mobility_a} is symmetric and positive definite, and eq~\eqref{approximate_effective_dynamic_equation_a_final}
is dissipative.

We can conclude that the auxiliary degrees of freedom $\bm{A}(t)$ (approximately) obey the Langevin type
equation \eqref{approximate_effective_dynamic_equation_a_final}.
There is no reversible term in eq~\eqref{approximate_effective_dynamic_equation_a_final},
in the same way as eq~\eqref{effective_dynamic_equation_phi_markovian}.
This would be natural because eq~\eqref{approximate_effective_dynamic_equation_a_final}
is the approximation for eq~\eqref{effective_dynamic_equation_phi} which has no reversible term.
Eq~\eqref{approximate_effective_dynamic_equation_a_final} is one of the main results of this work.
By combining the hypothetical transient potential $\Phi(\tilde{\bm{q}},t) \approx \check{\Phi}(\tilde{\bm{q}},\bm{A}(t))$ and eqs~\eqref{effective_dynamic_equation_q},
\eqref{effective_dynamic_equation_p}, and \eqref{approximate_effective_dynamic_equation_a_final},
we have the approximate Markovian dynamic equations for $\lbrace \bm{Q}(t), \bm{P}(t), \bm{A}(t) \rbrace$.
The non-Markovian dynamic equation \eqref{effective_dynamic_equation_phi} for the transient potential $\Phi(\tilde{\bm{q}},t)$
with infinite degrees of freedom is now reduced to the Markovian dynamic equation
\eqref{approximate_effective_dynamic_equation_a_final} for the auxiliary potential parameters $\bm{A}(t)$
with just $Z$ degrees of freedom.
We consider that the simplified approximate dynamic equation is still able to describe various complex dynamics
of the coarse-grained variables $\bm{Q}(t)$ and $\bm{P}(t)$.

We should recall that the approximation can be justified only if the hypothetical transient potential
reasonably mimics the full dynamics of the exact transient potential.
Therefore, the employed approximations may not be fully justified in some cases.
Nonetheless, this does not mean that the approximate dynamics model is not reasonable.
An illustrative example
is the coupled oscillator model in Sec.~\ref{coupled_oscillator_model}. 
As we showed, the transient
potential with the time-dependent potential parameter (eq~\eqref{transient_potential_coupled_oscillator}) perfectly reproduces
the dynamics of the exact transient potential for the coupled oscillator model.
The hypothetical transient potential with the potential parameters is exact in that case.
We consider that the result in this subsection justifies the use of the hypothetical transient potential with
potential parameters to
describe the coarse-grained dynamics. 
In Sec.~\ref{estimating_potential_parameters_from_transient_potential}, we will discuss 
one simple estimate method for the potential parameter
from a given transient potential.

\section{Discussions}
\label{discussions}

\subsection{Estimating Potential Parameters from Transient Potential}
\label{estimating_potential_parameters_from_transient_potential}

In Sec.~\ref{approximation_by_potential_parameters}, we showed that the dynamic equation for the transient potential
can be largely simplified by utilizing the potential parameters $\bm{A}(t)$.
However, we did not show methods to calculate the potential parameters from a given
transient potential. (There is essentially the same problem in the previous work\cite{Uneyama-2020}.)
If we want to directly compare the full dynamics of transient potential
and the approximate dynamics of the potential parameters, we will
need the method to estimate physically reasonable 
potential parameters $\bm{A}(t)$ from a given transient potential $\Phi(\tilde{\bm{q}},t)$, for
a given hypothetical transient potential $\check{\Phi}(\tilde{\bm{q}},\tilde{\bm{a}})$.
Here, we assume that we have the explicit functional form of $\check{\Phi}(\tilde{\bm{q}},\tilde{\bm{a}}$)
but the explicit values of the potential parameters are unknown (and thus the potential parameters are expressed as a dummy variable $\tilde{\bm{a}}$).
In molecular dynamics simulations, we have the full information
on the microscopic degrees of freedom. This means that we can directly evaluate
the transient potential $\Phi(\tilde{\bm{q}},t) = \hat{U}(\tilde{\bm{q}},\bm{\theta}(t))$.
Therefore, the method shown in this subsection may be interpreted as a method
to obtain the coarse-grained potential parameter $\bm{A}(t)$ from the microscopic degrees of freedom.

In principle, $\bm{A}(t)$ can depend on the history of $\Phi(\tilde{\bm{q}},t)$.
However, the analyses which involve the time-series of the potential parameters and
the transient potential will be too complex.
As a working hypothesis, here we simply assume that $\bm{A}(t)$ is determined
solely by the the coarse-grained position $\bm{Q}(t)$ and the transient potential $\Phi(\tilde{\bm{q}},t)$ at
the same $t$. (This would be interpreted as a sort of the Markov approximation which was utilized in Sec.~\ref{markov_approximation}.)
Under this assumption, it is sufficient for us to consider the estimate
method of the potential parameters $\bm{A}(t)$ from a
given transient potential $\Phi(\tilde{\bm{q}},t)$ at a certain time $t$.
Then the estimate method becomes essentially static.
Thus we simply ignore the time dependence in this subsection, and
describe the coarse-grained position, the transient potential, and the potential parameters as
$\bm{Q}$, $\Phi(\tilde{\bm{q}})$ and $\bm{A}$. Now what we should consider
is how to estimate $\bm{A}$ which gives the most reasonable approximation
for the transient potential, for given $\bm{Q}$, $\Phi(\tilde{\bm{q}})$, and $\check{\Phi}(\tilde{\bm{q}},\tilde{\bm{a}})$.

The potential parameters should be determined so that it reproduces the
original transient potential as accurate as possible. Intuitively, this
requirement is achieved if we minimize a sort of distance between the
original and hypothetical transient potentials.
Here we utilize the Kullback-Leibler divergence\cite{Kullback-Leibler-1951}
to measure how different two potentials are.
We employ the following Kullback-Leibler divergence for the probability distributions\cite{Shell-2008}:
\begin{equation}
 \label{kullback_leibler_divergence}
 \mathcal{K}(\tilde{\bm{a}})
  \equiv \int d\tilde{\bm{q}} \,
  \check{\Psi}_{\text{eq}}(\tilde{\bm{q}},\tilde{\bm{a}})
  \ln \frac{\check{\Psi}_{\text{eq}}(\tilde{\bm{q}},\tilde{\bm{a}})}
  {\Psi_{\text{eq}}(\tilde{\bm{q}})} ,
\end{equation}
where $\check{\Psi}_{\text{eq}}(\tilde{\bm{q}},\tilde{\bm{a}})$ is 
the equilibrium probability distribution under the hypothetical transient potential
$\check{\Phi}(\tilde{\bm{q}},\tilde{\bm{a}})$, and
 $\Psi_{\text{eq}}(\tilde{\bm{q}})$ is the equilibrium probability distribution under the transient potential $\Phi(\tilde{\bm{q}})$.
The Kullback-Leibler divergence satisfies $\mathcal{K}(\tilde{\bm{a}}) \ge 0$ and
it represents the discrepancy between two equilibrium distributions.
(It can be zero if and only if two equilibrium distributions are the same.)
We employ $\tilde{\bm{a}}$ which minimizes the Kullback-Leibler divergence
\eqref{kullback_leibler_divergence} as $\bm{A}$: $\partial \mathcal{K}(\tilde{\bm{a}}) / \partial \tilde{\bm{a}}|_{\tilde{\bm{a}} = \bm{A}} = 0$.
Although this method looks plausible, generally it is difficult to
calculate eq~\eqref{kullback_leibler_divergence} even numerically (since $\Phi_{\text{eq}}(\tilde{\bm{q}})$
and $\check{\Phi}_{\text{eq}}(\tilde{\bm{q}},\tilde{\bm{a}})$ (with fixed $\tilde{\bm{a}}$) are functions in the $M$-dimensional space).
Thus we need to introduce some approximations to minimize the Kullback-Leibler divergence
with a realistic calculation cost.

Although the Kullback-Leibler divergence \eqref{kullback_leibler_divergence} is
a functional of $\Phi(\tilde{\bm{q}})$ and $\check{\Phi}(\tilde{\bm{q}},\tilde{\bm{a}})$,
we may not need the full information of the transient potentials.
We limit ourselves to consider the contributions around 
a given value of the coarse-grained positions $\bm{Q}$.
However, in general, the transient potential may not be locally stable.
We introduce a simple trick here. We virtually add
a trap potential
to the system:
\begin{equation} 
\Phi_{\text{trap}}(\tilde{\bm{q}}) \equiv \frac{1}{2} 
 (\tilde{\bm{q}} - \bm{Q}) \cdot \bm{C}_{\text{trap}} \cdot
  (\tilde{\bm{q}} - \bm{Q}),
\end{equation}
where $\bm{C}_{\text{trap}}$ is
a symmetric positive definite tensor. We employ the equilibrium probability distributions
under the transient potentials and the trap potential,
$\Psi_{\text{eq}}(\tilde{\bm{q}}) \propto \exp[-\Phi(\tilde{\bm{q}}) / k_{B} T - \Phi_{\text{trap}}(\tilde{\bm{q}}) / k_{B} T]$
and
$\check{\Psi}_{\text{eq}}(\tilde{\bm{q}}) \propto \exp[-\check{\Phi}(\tilde{\bm{q}},\tilde{\bm{a}}) / k_{B} T - \Phi_{\text{trap}}(\tilde{\bm{q}}) / k_{B} T]$,
to calculate the Kullback-Leibler divergence.
Due to the trap potential, the equilibrium
probability distributions should be stably localized around $\tilde{\bm{q}} \approx \bm{Q}$.
Thus we expand transient potentials 
into the power series of
$\Delta \tilde{\bm{q}}(\bm{Q}) \equiv \tilde{\bm{q}} - \bm{Q}$ and approximate them as
\begin{align}
 \label{transient_potential_harmonic_approximation}
 \Phi(\tilde{\bm{q}}) & \approx 
 \Phi(\bm{Q}) - \bm{F}(\bm{Q}) \cdot \Delta\tilde{\bm{q}}(\bm{Q})
 + \frac{1}{2} \Delta\tilde{\bm{q}}(\bm{Q}) \cdot \bm{C}(\bm{Q}) \cdot \Delta\tilde{\bm{q}}(\bm{Q}), \\
 \label{hypothetical_transient_potential_harmonic_approximation}
 \check{\Phi}(\tilde{\bm{q}},\tilde{\bm{a}}) & \approx 
 \check{\Phi}(\bm{Q},\tilde{\bm{a}}) - \check{\bm{F}}(\bm{Q},\tilde{\bm{a}}) \cdot \Delta\tilde{\bm{q}}(\bm{Q})
 + \frac{1}{2} \Delta\tilde{\bm{q}}(\bm{Q}) \cdot \check{\bm{C}}(\bm{Q},\tilde{\bm{a}}) \cdot \Delta\tilde{\bm{q}}(\bm{Q}).
\end{align}
Here, $\bm{F}(\bm{Q}) \equiv - \partial \Phi(\bm{Q}) / \partial \bm{Q}$,
$\check{\bm{F}}(\bm{Q},\tilde{\bm{a}}) \equiv - \partial \check{\Phi}(\bm{Q},\tilde{\bm{a}}) / \partial \bm{Q}$,
$\bm{C}(\bm{Q}) \equiv \partial^{2} \Phi(\bm{Q}) / \partial \bm{Q} \partial \bm{Q}$,
and $\check{\bm{C}}(\bm{Q},\tilde{\bm{a}}) \equiv \partial^{2} \check{\Phi}(\bm{Q},\tilde{\bm{a}}) / \partial \bm{Q} \partial \bm{Q}$ are expansion coefficients.
(In the remaining part of this subsection, for simplicity, we do not explicitly describe arguments for these expansion coefficients.
The expansion coefficients of the original transient potential are functions of $\bm{Q}$,
whereas those of the hypothetical transient potential are functions of $\bm{Q}$ and $\tilde{\bm{a}}$.)
With these approximations, the Kullback-Leibler divergence \eqref{kullback_leibler_divergence}
can be (approximately) simplified as
\begin{equation}
 \label{kullback_leibler_divergence_approx}
 \mathcal{K}(\tilde{\bm{a}})
 \approx \frac{1}{4} \mathop{\mathrm{tr}} (\bm{G} \cdot \Delta \check{\bm{C}})^{2}
 + \frac{1}{2 k_{B} T} (\Delta \check{\bm{F}} - \bm{F} \cdot \bm{G} \cdot \Delta \check{\bm{C}})
 \cdot \bm{G} \cdot  (\Delta \check{\bm{F}} - \Delta \check{\bm{C}} \cdot \bm{G} \cdot \bm{F} ),
\end{equation}
where $\bm{G} \equiv (\bm{C}_{\text{trap}}+ \bm{C})^{-1}$, $\Delta \check{\bm{C}} \equiv \check{\bm{C}} - \bm{C}$,
and $\Delta \check{\bm{F}} \equiv \check{\bm{F}} - \bm{F}$. See Appendix~\ref{calculation_of_kullback_leibler_divergence} for detailed calculations.
Eq~\eqref{kullback_leibler_divergence_approx} can be numerically evaluated with reasonable calculation costs
($\bm{G}^{-1}$ is expected to be sparse in most cases). Thus we can numerically find $\tilde{\bm{a}} = \bm{A}$
which minimizes the Kullback-Leibler divergence, and thus we can 
estimate the potential parameter $\bm{A}$ from
the transient potential $\Phi(\tilde{\bm{q}})$.
If the trap potential is strong, we may introduce an additional approximation for $\bm{G}$ as $\bm{G} \approx {\bm{C}}_{\text{trap}}^{-1} - \bm{C}_{\text{trap}}^{-1} \cdot \bm{C} \cdot \bm{C}_{\text{trap}}^{-1}$.
In addition, if we take $\bm{C}_{\text{trap}}$ to be diagonal, $\bm{G}$ can be calculated
straightforwardly.

Here we show some simple examples for the estimate of the potential parameters from transient potentials.
One simple example is the transient potential
for the coupled oscillator model in Sec.~\ref{coupled_oscillator_model}.
If we interpret the transient potential \eqref{transient_potential_coupled_oscillator} as the
hypothetical transient potential, $\kappa_{\text{eff}}$ and $A$ are the potential parameters.
The minimization of the Kullback-Leibler divergence gives eq~\eqref{potential_parameters_coupled_oscillator}.
(In this case, the Kullback-Leibler divergence becomes zero if we set $\kappa_{\text{eff}}$ and $A$
as eq~\eqref{potential_parameters_coupled_oscillator}, and
the hypothetical transient potential perfectly reproduces the original transient potential.)
Thus we find that our method works reasonably, at least in this simple case.

Another simple example is the transient potential for
a single tagged particle in a supercooled liquid. At the short-time scale, the tagged
particle will feel the cage potential formed by surrounding particles.
This cage potential can be interpreted as a transient potential\cite{Hachiya-Uneyama-Kaneko-Akimoto-2019}.
The center of the cage can move or the potential may be destroyed and then newly created.
We employ the following hypothetical transient potential:
\begin{equation}
 \check{\Phi}(\tilde{\bm{q}},\tilde{\bm{a}},\tilde{s}) = \frac{1}{2} \kappa \tilde{s} 
  (\tilde{\bm{q}} - \tilde{\bm{a}})^{2} ,
\end{equation}
where $\kappa$ is a positive constant, $\tilde{\bm{a}}$ and $\tilde{s}$ are
dummy potential parameters which represent the cage center position and
the cage state. We assume
that $\tilde{s}$ can take either $0$ or $1$. ($\tilde{s} = 0$ corresponds to the
free state, and $\tilde{s} = 1$ corresponds to the
trapped state.) The Kullback-Leibler divergence becomes
\begin{equation}
 \label{kullback_leibler_divergence_for_cage_potential}
\begin{split}
 \mathcal{K}(\tilde{\bm{a}},\tilde{s})
 & \approx \frac{1}{4} \mathop{\mathrm{tr}}
 \left[\bm{G} \cdot (\kappa \tilde{s} \bm{1} - \bm{C} )\right]^{2}
 + \frac{1}{2 k_{B} T}
 \left[ \kappa \tilde{s} (\tilde{\bm{a}} - \bm{Q}) - \bm{F} - \bm{F} \cdot \bm{G} \cdot (\kappa \tilde{s} \bm{1} - \bm{C} ) \right] \\
 & \qquad \cdot \bm{G} \cdot 
  \left[ \kappa \tilde{s} (\tilde{\bm{a}}- \bm{Q}) - \bm{F} - (\kappa \tilde{s} \bm{1} - \bm{C} ) \cdot \bm{G} \cdot \bm{F} \right].
\end{split}
\end{equation}
Note that, $\bm{F} = - \partial \Phi(\bm{Q}) / \partial \bm{Q}$ and $\bm{C} = \partial^{2} \Phi(\bm{Q}) / \partial \bm{Q} \partial \bm{Q}$
can be calculated straightforwardly if the transient potential is given.

We calculate $\tilde{\bm{a}}$ and $\tilde{s}$ which minimizes the Kullback-Leibler divergence
and employ them as the potential parameters $\bm{A}$ and $S$.
If $\tilde{s} = 0$, eq~\eqref{kullback_leibler_divergence_for_cage_potential}
becomes independent of $\tilde{\bm{a}}$:
\begin{equation}
 \label{kullback_leibler_divergence_for_cage_potential_s0}
 \mathcal{K}(\tilde{\bm{a}},0)
  \approx \frac{1}{4} \mathop{\mathrm{tr}}
 (\bm{G} \cdot \bm{C} )^{2}
 + \frac{1}{2 k_{B} T}
 \bm{F} \cdot (\bm{1} - \bm{G} \cdot \bm{C})  \cdot \bm{G}
 \cdot (\bm{1} - \bm{C} \cdot \bm{G}) \cdot \bm{F}.
\end{equation}
On the other hand, if $\tilde{s} = 1$, eq~\eqref{kullback_leibler_divergence_for_cage_potential}
depends on $\tilde{\bm{a}}$. Thus we should minimize it with respect to $\tilde{\bm{a}}$.
We have the following value as the minimized Kullback-Leibler divergence:
\begin{equation}
 \label{kullback_leibler_divergence_for_cage_potential_s1}
 \mathcal{K}(\bm{A}_{\text{mp}},1)
  \approx \frac{1}{4} \mathop{\mathrm{tr}}
 \left[\bm{G} \cdot (\kappa \bm{1} - \bm{C} )\right]^{2}
\end{equation}
with $\bm{A}_{\text{mp}} \equiv \bm{Q} + [\bm{F} + (\kappa \bm{1} - \bm{C} ) \cdot \bm{G} \cdot \bm{F}] / \kappa$
being the most probable value of $\tilde{\bm{a}}$.
By comparing the Kullback-Leibler divergence values by eqs~\eqref{kullback_leibler_divergence_for_cage_potential_s0} and
\eqref{kullback_leibler_divergence_for_cage_potential_s1}, we can determine
$S$ for a given transient potential $\Phi(\tilde{\bm{q}})$. 
If $S = 0$, we do not need $\bm{A}$ because the hypothetical transient potential becomes independent of $\bm{A}$.
If $S = 1$, $\bm{A}$ can be determined straightforwardly
as $\bm{A} = \bm{A}_{\text{mp}}$.

By utilizing the method shown above, we will be able to
construct the time-dependent potential parameters from the
time-dependent transient potential. This method would be utilized to
directly extract the information of the transient potential from such
as the molecular dynamics simulation data. In the molecular dynamics simulations,
the interaction potentials between particles are generally nonlinear.
However, in our estimate method, only the local information around the
current coarse-grained positions is utilized (eq~\eqref{transient_potential_harmonic_approximation}). Thus even if
the potentials are strongly nonlinear, our method based on eq~\eqref{kullback_leibler_divergence_approx}
will work as long
as the expansion forms (with the expansion coefficients $\bm{F}$ and $\bm{C}$)
are reasonable.
Conversely, if the potentials are not smooth (such as the hard-sphere potential),
our method will not work.
The overall accuracy of the approximation
by the potential parameters would be examined by calculating the
Kullback-Leibler divergence. If we have several different functional forms
for the hypothetical transient potential as possible candidates,
we can judge which is the best from the Kullback-Leibler divergence.

\subsection{Comparison with Generalized Langevin Equation}
\label{comparison_with_generalized_langevin_equation}

The coarse-grained dynamic equations with the transient potential derived in
this work is just one possible coarse-graining model, and we can construct
other coarse-grained models for the same microscopic system.
In this subsection, we compare our model with the standard coarse-grained dynamics model:
the generalized Langevin equation.

As we mentioned, the projection operator method is a standard technique to
construct the coarse-grained dynamic equations.
In the conventional coarse-graining method, we consider only $\bm{Q}(t)$ and $\bm{P}(t)$ as the coarse-graining degrees of freedom,
and do not employ the transient potential.
Then we have the following generalized Langevin equations for $\bm{Q}(t)$ and $\bm{P}(t)$\cite{Kawasaki-1973,Dengler-2016}:
\begin{align}
 \label{effective_dynamic_equation_q_conventional}
 \frac{d\bm{Q}(t)}{dt} & = \bm{M}^{-1} \cdot \bm{P}(t),  \\
 \begin{split}
 \label{effective_dynamic_equation_p_conventional}
 \frac{d\bm{P}(t)}{dt} & = - \frac{\partial \bar{\mathcal{F}}(\bm{Q}(t))}{\partial \bm{Q}(t)} 
  - \int_{-\infty}^{t} dt' \, \bar{\bm{K}}'(\bm{Q}(t'),\bm{P}(t'),t - t') \cdot \bm{M}^{-1} \cdot \bm{P}(t') \\
  & \qquad + k_{B} T \int_{-\infty}^{t} dt' \, \frac{\partial}{\partial \bm{P}(t')} \cdot \bar{\bm{K}}'^{\mathrm{T}}(\bm{Q}(t'),\bm{P}(t'),t - t')
 + \bm{\xi}'(t) ,
 \end{split}
\end{align}
where $\hat{\bm{K}}'(\bm{Q},\bm{P},t)$ is the memory kernel tensor and $\bm{\xi}'(t)$ is the noise vector.
They satisfy the fluctuation-dissipation relation:
\begin{equation}
 \langle \bm{\xi}'(t) \rangle_{\text{eq}} = 0, \qquad
 \langle \bm{\xi}'(t) \bm{\xi}'(t') \rangle_{\text{eq}} = k_{B} T \langle \bar{\bm{K}}'(\bm{Q},\bm{P},|t - t'|)\rangle_{\text{eq}} .
\end{equation}

Eq~\eqref{effective_dynamic_equation_q_conventional} is the same as eq~\eqref{effective_dynamic_equation_q}
but eq~\eqref{effective_dynamic_equation_p_conventional} is much different from
eq~\eqref{effective_dynamic_equation_p}. This is because the transient potential is not included
in this case.
In eq~\eqref{effective_dynamic_equation_p_conventional}, the time-evolution of $\bm{P}(t)$ is primarily governed by 
the force by the free energy $- \partial \bar{\mathcal{F}}(\bm{Q}(t)) / \partial \bm{Q}(t)$,
and the eliminated fast degrees of freedom work as the friction (with the memory effect)
and the noise. From the viewpoint of the transient potential,
the time-evolution of $\bm{P}(t)$ is governed by the force by the transient potential,
$- \partial \Phi(\bm{Q}(t),t) / \partial \bm{Q}(t)$ (eq~\eqref{effective_dynamic_equation_p}).
We decompose this force into the average
$- \partial \bar{\mathcal{F}}(\bm{Q}(t)) / \partial \bm{Q}(t)$ 
and the time-dependent fluctuation $- \partial [\Phi(\bm{Q}(t),t) - \bar{\mathcal{F}}(\bm{Q}(t))] / \partial \bm{Q}(t)$.
We may interpret that the fluctuation part,
$- \partial [\Phi(\bm{Q}(t),t) - \bar{\mathcal{F}}(\bm{Q}(t))] / \partial \bm{Q}(t)$, works
as the friction and noise terms {\hl in the generalized Langevin equation}. 
Intuitively, we expect that eq~\eqref{effective_dynamic_equation_p_conventional} is
reproduced if we eliminate the transient potential $\Phi(\tilde{\bm{q}},t)$ from
eqs~\eqref{effective_dynamic_equation_p} and \eqref{effective_dynamic_equation_phi}.
In this sense, our dynamic equations with the transient potential
is less coarse-grained than the generalized Langevin equations.

Here, it would be fair to mention that eqs~\eqref{effective_dynamic_equation_q_conventional} and \eqref{effective_dynamic_equation_p_conventional}
are formally exact. There are several different ways to rewrite the microscopic
dynamic equations into coarse-grained dynamic equations (as shown in Sec.~\ref{coupled_oscillator_model}). The resulting effective dynamic equations depend on
various factors such as
the choice of the degrees of freedom, and the operator identity for the decomposition\cite{Chaturvedi-Shibata-1979,Uchiyama-Shibata-1999}.
Our coarse-grained dynamics model in Sec~\ref{theory} should be interpreted as just one possible candidate.
If one wants to describe the coarse-grained dynamics without any additional degrees of
freedom, the generalized Langevin equation would be suitable. Then the memory kernel is required
to reproduce the memory effect. If one does not want to use the memory kernel, yet still
want to incorporate the memory effect,
the transient potential (with the Markov and other approximations) would be utilized instead.
If there are some other coarse-grained dynamics models,
of course they would be also possible candidates.

\subsection{Comparison with Langevin Equation with Transient Potential}
\label{comparison_with_langevin_equation_with_transient_potential}

In the previous work\cite{Uneyama-2020}, the dynamic equation for $\bm{Q}(t)$ was expressed as
the LETP. This is in
contrast to the result in this work. We have obtained
the Hamiltonian-like dynamics for $\bm{Q}(t)$ and $\bm{P}(t)$
(eqs~\eqref{effective_dynamic_equation_q} and \eqref{effective_dynamic_equation_p}).
Thus, the dynamic equations obtained in this work may seem to be inconsistent
with the LETP. In this subsection, we show that, our model can
reduce to the LETP under some approximations.

We assume that the the potential parameters and transient potential can
be decomposed into relatively fast and slow parts:
$\check{\Phi}(\tilde{\bm{q}},\bm{A}(t)) = \check{\Phi}^{(\text{f})}(\tilde{\bm{q}},\bm{A}^{(\text{f})}(t))
+ \check{\Phi}^{(\text{s})}(\tilde{\bm{q}},\bm{A}^{(\text{s})}(t))$. Here, the superscripts ``(f)'' and ``(s)''
represent the fast and slow parts, respectively.
We also assume that the effective potential can be separated into two parts, in the same manner: $\check{\Upsilon}(\bm{A})
= \check{\Upsilon}^{(\text{f})}(\bm{A}^{(\text{f})}) + \check{\Upsilon}^{(\text{s})}(\bm{A}^{(\text{s})})$.
We consider the hypothetical case where the dynamic equations for fast and slow auxiliary degrees of freedom
are not kinetically coupled.
The fast auxiliary degrees of freedom are expected to represent the fluctuation around the 
locally stable position for $\bm{Q}(t)$. Thus we assume that $\bm{A}^{(\text{f})}(t)$ has the 
same dimension as $\bm{Q}(t)$. ($\bm{A}^{(\text{f})}(t)$ is an $M$-dimensional vector
whereas $\bm{A}^{(\text{s})}(t)$ is a $(Z - M)$-dimensional vector.)
We expect that the fluctuation of $\bm{A}^{(\text{f})}(t)$ around the locally stable position
are not large, and thus a simple harmonic potential would be sufficient to
approximately express the fast part of the transient potential.
If we eliminate the fast auxiliary degrees of freedom, we will have the
coarse-grained dynamic equations for relatively slow degrees of freedom.
This corresponds to the coarse-graining for
our coarse-grained dynamics model.

Based on the discussions above, we model the dynamics of the fast auxiliary degrees of freedom as the simple Gaussian
stochastic process in the harmonic potential:
\begin{align}
 \label{langevin_equation_a_fast}
 \frac{d\bm{A}^{(\text{f})}(t)}{dt} 
 & = - \bm{L}^{(\text{f})} \cdot \frac{\partial  \check{\Phi}^{(\text{f})}(\bm{Q}(t),\bm{A}^{(\text{f})}(t)) }{\partial \bm{A}^{(\text{f})}(t)} + \sqrt{2 k_{B} T}
 \bm{B}^{(\text{f})} \cdot \bm{W}^{(\text{f})}(t), \\
 \check{\Phi}^{(\text{f})}(\bm{Q},\bm{A}^{(\text{f})}) 
 & = \frac{1}{2} (\bm{Q} - \bm{A}^{(\text{f})}) \cdot \bm{\kappa} \cdot (\bm{Q} - \bm{A}^{(\text{f})}) .
\end{align}
Here, $\bm{\kappa}$ is a symmetric positive-definite second order tensor, and is
assumed to be independent of $\bm{Q}(t)$. 
We have assumed that $\check{\Upsilon}^{(\text{f})}(\bm{A}^{(\text{f})}) = (\text{const.})$ and
the contribution of $\check{\Upsilon}^{(\text{f})}(\bm{A}^{(\text{f})})$ disappears in eq~\eqref{langevin_equation_a_fast}.
$\bm{L}^{(\text{f})}$ is the mobility tensor and is approximated to be constant.
$\bm{B}^{(\text{f})}$ is the noise coefficient tensor which is defined via
$\bm{L}^{(\text{f})} = \bm{B}^{(\text{f})} \cdot \bm{B}^{(\text{f}) \mathrm{T}}$.
Also, we assume that the noise vector $\bm{W}^{(\text{f})}(t)$ is the Gaussian white noise vector.
Its first and second moments are
$ \langle \bm{W}^{(\text{f})}(t) \rangle_{\text{eq}} = 0$ and
$\langle \bm{W}^{(\text{f})}(t) \bm{W}^{(\text{f})}(t') \rangle_{\text{eq}} = \bm{1} \delta(t - t')$.
For the dynamics of the slow auxiliary degrees of freedom, we 
employ the following Langevin type equation which has the same form as eq~\eqref{approximate_effective_dynamic_equation_a_final}:
\begin{equation}
 \label{langevin_equation_a_slow}
\begin{split}
 \frac{d \bm{A}^{(\text{s})}(t)}{d t} 
 & \approx - 
  \check{\bm{L}}^{(\text{s})}(\bm{Q}(t),\bm{P}(t),\bm{A}^{(\text{s})}(t)) \cdot
 \left[ 
 \frac{\partial \check{\Phi}^{(\text{s})}(\bm{Q}(t),\bm{A}^{(\text{s})}(t))}{\partial \bm{A}^{(\text{s})}(t)} 
 + \frac{\partial \check{\Upsilon}^{(\text{s})}(\bm{A}^{(\text{s})}(t))}{\partial \bm{A}^{(\text{s})}(t)} 
 \right]\\
 & \qquad + k_{B} T \frac{\partial}{\partial \bm{A}^{(\text{s})}(t)} \cdot 
 \check{\bm{L}}^{(\text{s})}(\bm{Q}(t),\bm{P}(t),\bm{A}^{(\text{s})}(t)) \\
 & \qquad + \sqrt{2 k_{B} T}
 \check{\bm{B}}^{(\text{s})}(\bm{Q}(t),\bm{P}(t),\bm{A}^{(\text{s})}(t)) \cdot
 \bm{W}^{(\text{s})}(t).
\end{split}
\end{equation}
Here, $\check{\bm{L}}^{(\text{s})}(\bm{Q},\bm{P},\bm{A}^{(\text{s})})$
and $\check{\bm{B}}^{(\text{s})}(\bm{Q},\bm{P},\bm{A}^{(\text{s})})$
are the mobility and noise tensors, respectively, and they
are assumed to be independent of $\bm{A}^{(\text{f})}(t)$.
$\bm{W}^{(\text{s})}(t)$ is the noise vector and its
first and second moments are given as
$ \langle \bm{W}^{(\text{s})}(t) \rangle_{\text{eq}} = 0$ and
$\langle \bm{W}^{(\text{s})}(t) \bm{W}^{(\text{s})}(t') \rangle_{\text{eq}} = \bm{1} \delta(t - t')$.
As before, the noise vector $\bm{W}^{(\text{s})}(t)$ can be non-Gaussian.

Eq~\eqref{langevin_equation_a_fast} can be easily solved to give
\begin{equation}
 \bm{A}^{(\text{f})}(t) = \bm{Q}(t) +
  \int_{-\infty}^{t} dt' \, \bm{\kappa}^{-1} \cdot \bm{K}^{(\text{f})}(t - t') \cdot 
  \left[ - \frac{d\bm{Q}(t')}{dt'} + \sqrt{2 k_{B} T} \bm{B}^{(\text{f})} \cdot \bm{W}^{(\text{f})}(t')  \right],
\end{equation}
with $\bm{K}^{(\text{f})}(t) \equiv \bm{\kappa} \cdot
\exp( -t \bm{L}^{(\text{f})} \cdot  \bm{\kappa})$
being the memory kernel tensor.
Then the dynamic equation for $\bm{P}(t)$ (eq~\eqref{effective_dynamic_equation_p}) can be rewritten as
\begin{equation}
 \label{effective_dynamic_equation_p_decomposed}
  \begin{split}
  \frac{d\bm{P}(t)}{dt} 
   & =
 - \frac{\partial \check{\Phi}^{(\text{s})}(\bm{Q}(t),\bm{A}^{(\text{s})}(t))}{\partial \bm{Q}(t)} \\
   & \qquad +  \int_{-\infty}^{t} dt' \, \bm{K}^{(\text{f})}(t - t') \cdot 
 \left[ - \frac{d\bm{Q}(t')}{dt'} + \sqrt{2 k_{B} T} \bm{B}^{(\text{f})} \cdot \bm{W}^{(\text{f})}(t')   \right].
  \end{split}
\end{equation}
Eq~\eqref{effective_dynamic_equation_p_decomposed}
can be interpreted as the generalized Langevin equation with the transient
potential $\check{\Phi}^{(\text{s})}(\tilde{\bm{q}},\bm{A}^{(\text{s})})$.
If the fast auxiliary degrees of freedom relax much faster than
the other degrees of freedom, we can approximate the memory kernel as:
\begin{equation}
 \int_{-\infty}^{t} dt' \, \bm{K}^{(\text{f})}(t - t') \cdot \bm{f}(t')
  \approx \left[ \int_{-\infty}^{t} dt' \, \bm{K}^{(\text{f})}(t - t') \right]
\cdot \bm{f}(t)
\equiv \bm{L}^{(\text{f})\, -1} \cdot \bm{f}(t),
\end{equation}
where $\bm{f}(t)$ is an arbitrary time-dependent vector which varies slowly
compared with $\bm{A}^{(\text{f})}(t)$.
Then, we have the following dynamic equation for the coarse-grained positions $\bm{Q}(t)$:
\begin{equation}
 \label{underdamped_letp_from_transient_potential_dynamics}
 \bm{M} \cdot \frac{d^{2}\bm{Q}(t)}{dt^{2}} \approx
 - \frac{\partial \check{\Phi}^{(\text{s})}(\bm{Q}(t),\bm{A}^{(\text{s})}(t))}{\partial \bm{Q}(t)}
 - \bm{L}^{(\text{f})\, -1} \cdot \frac{d\bm{Q}(t)}{dt}
 + \sqrt{2 k_{B} T} \bm{L}^{(\text{f})\, -1} \cdot \bm{B}^{(\text{f})} \cdot \bm{W}^{(\text{f})}(t),
\end{equation}
where we have used $\bm{P}(t) = \bm{M} \cdot d\bm{Q}(t) / dt$.
We can interpret $\bm{L}^{(\text{f})\, -1}$ as the
effective friction coefficient tensor for the coarse-grained positions $\bm{Q}(t)$.
Thus we have the underdamped LETP with the transient potential $\check{\Phi}^{(\text{s})}(\tilde{\bm{q}},\bm{A}^{(\text{s})}(t))$ as the approximation of our dynamics model.

If the inertia effect is weak,
we can take the zero-mass limit ($\bm{M} \to 0$) as a reasonable approximation. Then the dynamic
equation \eqref{underdamped_letp_from_transient_potential_dynamics} can be further simplified as:
\begin{equation}
 \label{overdamped_letp_from_transient_potential_dynamics}
 \frac{d\bm{Q}(t)}{dt} \approx
 - \bm{L}^{(\text{f})} \cdot \frac{\partial \check{\Phi}^{(\text{s})}(\bm{Q}(t),\bm{A}^{(\text{s})}(t))}{\partial \bm{Q}(t)}
 + \sqrt{2 k_{B} T} \bm{B}^{(\text{f})}\cdot \bm{W}^{(\text{f})}(t). 
\end{equation}
Eq~\eqref{overdamped_letp_from_transient_potential_dynamics} has exactly the same form as the overdamped LETP in Ref.~\cite{Uneyama-2020}.
Eqs~\eqref{overdamped_letp_from_transient_potential_dynamics} and \eqref{langevin_equation_a_slow}
describe the dynamics of the reduced coarse-grained variables $\bm{Q}(t)$ and $\bm{A}^{(\text{s})}(t)$ in a closed form.
Therefore, we conclude that, the LETP can be successfully obtained as a coarse-grained
dynamic equation, starting from the microscopic Hamiltonian dynamics.
We should recall that the LETP \eqref{overdamped_letp_from_transient_potential_dynamics}
is not exact. To derive eq~\eqref{overdamped_letp_from_transient_potential_dynamics}, we have
utilized various approximations: the Markovian approximation, the hypothetical
transient potential with potential parameters, the decomposition of the potential
parameters into the fast and slow parts, the harmonic transient potential and simple Langevin dynamics for the decomposed fast part,
and the zero-mass limit.
Some of the approximations are rough and may not be fully justified.
Nonetheless, we consider that the derivation shown above would be
informative when we use the LETP as a
coarse-grained dynamics model.

If we accept approximations involved in the derivation of the LETP,
we find that
the (overdamped) LETP can be obtained in two different ways.
One way is to directly utilize the projection operator to the microscopic Hamiltonian dynamics (as shown in this work).
The Hamiltonian dynamics is first reduced to the Hamiltonian-like dynamics
with the transient potential. Then by eliminating the relatively fast 
auxiliary degrees of freedom with some approximations,
we have the LETP as the further coarse-grained dynamic equation.
Another way is to derive the Langevin equation and then perform the coarse-graining for it. By using the
projection operator method, we have the generalized Langevin equation from
the microscopic Hamiltonian dynamics
(as shown in Sec.~\ref{comparison_with_generalized_langevin_equation}).
With some approximations, the generalized Langevin equation can reduce
to the overdamped Langevin equation. Then, by interpreting the overdamped Langevin equation as
a microscopic model, and eliminating relatively fast
degrees of freedom, we have the LETP (as shown in the previous work\cite{Uneyama-2020}).
These two ways are not identical but both of them can reproduce the LETP.
We expect that this is because the LETP is a physically natural coarse-grained dynamic
equation model.

We should comment that $\check{\Upsilon}^{(\text{s})}(\bm{A}^{(\text{s})})$ (in eq~\eqref{langevin_equation_a_slow})
do not appear in the previous work. One may consider that the result of this work
is not fully consistent with the previous work. This is due to the difference
between the transient potentials in this and previous works.
We can interpret that
$\check{\Phi}_{\text{eff}}(\tilde{\bm{q}},\bm{A}^{(\text{s})}) = \check{\Phi}^{(\text{s})}(\tilde{\bm{q}},\bm{A}^{(\text{s})}) + \check{\Upsilon}^{(\text{s})}(\bm{A}^{(\text{s})})$
corresponds to the transient potential in the previous work.
In the LETP, as we can observe in eq~\eqref{overdamped_letp_from_transient_potential_dynamics}, only the derivatives of the
transient potentials with respect to $\tilde{\bm{q}}$ are included. Since
$\partial \check{\Phi}_{\text{eff}}(\tilde{\bm{q}},\bm{A}^{(\text{s})})  / \partial \tilde{\bm{q}}
= \partial \check{\Phi}^{(\text{s})}(\tilde{\bm{q}},\bm{A}^{(\text{s})}) / \partial \tilde{\bm{q}}$, the LETP is
essentially not changed if we employ $\check{\Phi}_{\text{eff}}(\tilde{\bm{q}},\bm{A}^{(\text{s})})$
as the transient potential.
Also, the free energy functional for $\lbrace \bm{Q},\bm{P},\bm{A}^{(\text{s})} \rbrace$
satisfies
$\partial \check{\Omega}^{\text{(s)}}(\bm{Q},\bm{P},\bm{A}^{(\text{s})}) / \partial \bm{A}^{(\text{s})}
= \partial \check{\Phi}_{\text{eff}}(\bm{Q},\bm{A}^{(\text{s})})  / \partial \bm{A}^{(\text{s})}$.
Thus the dynamic equation for $\bm{A}^{(\text{s})}(t)$ (eq~\eqref{langevin_equation_a_slow})
can be described 
only with the transient potential $\check{\Phi}_{\text{eff}}(\bm{Q},\bm{A}^{(\text{s})})$.
We consider that the results in this work is consistent with the previous work.

\section{Conclusions}
\label{conclusions}

In this work, we showed that the dynamic equations with
the transient potential can be derived from the microscopic Hamiltonian dynamics.
We showed that the coupled oscillator model can be rewritten as the dynamics model
with the transient potential (Sec.~\ref{coupled_oscillator_model}).
The dynamic equations with the transient potential (eqs~\eqref{dynamic_equation_coupled_oscillator_transient_potential_q}-\eqref{dynamic_equation_coupled_oscillator_transient_potential_a})
are exact, and can be employed as an alternative to the generalized Langevin type equations (eqs~\eqref{gle_coupled_oscillator_q} and \eqref{gle_coupled_oscillator_p}).
This implies
that the dynamics model with the transient potential can be used as a
coarse-grained dynamics model in various systems.

Then we derived the dynamic equations with the transient potential,
starting from the microscopic Hamiltonian dynamics for a more general system (Sec.~\ref{theory}).
To derive the coarse-grained dynamic equation with the transient potential,
we introduced the projection operator for the coarse-grained variables and
the transient potential (eq~\eqref{projection_operator_p}).
We derived the dynamic equations for the coarse-grained
positions and momenta, $\bm{Q}(t)$ and $\bm{P}(t)$, and the transient potential $\Phi(\tilde{\bm{q}},t)$.
The dynamic equations for the positions and momenta have almost the same forms as 
the canonical equations (eqs~\eqref{effective_dynamic_equation_q} and \eqref{effective_dynamic_equation_p}),
whereas the dynamic equation for the transient potential (eq~\eqref{effective_dynamic_equation_phi})
is the stochastic partial differential equation with the memory kernel.
The transient potential fluctuates around the free energy $\bar{\mathcal{F}}(\tilde{\bm{q}})$.
The dynamic equation for the transient potential is exact
but too complex and not suitable for practical purposes.

We introduced several approximations to make the dynamic equation
for the transient potential simple and tractable.
By employing
the Markov approximation, we ignored the memory effect and simplified the dynamic
equation for the transient potential.
The approximate Markovian dynamic equation (eq~\eqref{effective_dynamic_equation_phi_markovian}) is,
however, still not simple. In order to further simplify the dynamic equation, we
introduced a hypothetical transient potential with
the time-dependent potential parameters $\bm{A}(t)$.
With some additional approximations, the dynamic equation for $\Phi(\tilde{\bm{q}},t)$ finally reduced
to the Langevin type dynamic equation for $\bm{A}(t)$ (eq~\eqref{approximate_effective_dynamic_equation_a_final}).
For the estimate of the potential parameters from the transient potential,
we proposed a simple method based on the Kullback-Leibler divergence and
showed that it works reasonably for some simple cases.
Thus we conclude that 
we can derive the relatively simple dynamics model with the transient potential
starting from the microscopic Hamiltonian dynamics, although the employed approximations may not be fully justified.
Moreover, by employing further approximations, our dynamics model reduces to
the LETP after the further coarse-graining. Therefore we consider that 
this work justifies the use of the LETP as the coarse-grained
model based on the microscopic Hamiltonian dynamics.

We expect that this work supports the use of the transient potential
to model mesoscopic coarse-grained dynamics.
Now we consider that the concept of the transient potential is not
phenomenological, rather based on the microscopic Hamiltonian dynamics
and a statistical mechanical basis.
The LETP in Ref.~\cite{Uneyama-2020} can be derived from the microscopic
Hamiltonian dynamics with some approximations.
The coarse-grained dynamics models with the transient potentials can
be employed as statistical-mechanically appropriate models.
Even if we accept the concept
of the transient potential dynamics, whether it is really efficient and
useful for mesoscopic coarse-grained modeling is still not fully clear. We will need to apply
our method to some simple mesoscopic dynamics to validate it.
For example, the coarse-graining of interacting many particle systems
(used in the molecular dynamics simulations) into
coarse-grained particles interacting via the transient potentials
is an interesting and important target.
If we apply our method to interacting many polymer systems, we will be
able to construct the coarse-grained dynamics for such as centers of mass of polymers
and end-to-end vectors. It will justify (or reconstruct)
the RaPiD model from the underlying microscopic dynamics model, and is
an interesting future work. Another interesting future work
is to construct several phenomenological dynamics models in the framework
proposed in this work. Although our derivation is rather formal,
we consider it is informative when we construct phenomenological
coarse-grained models.

\section*{Acknowledgment}

The author thanks an anonymous reviewer of Ref~\cite{Uneyama-2020} for
critically pointing that the derivation in Ref~\cite{Uneyama-2020}
lacks the connection to the microscopic dynamics.
The author also thanks Dr Takenobu Nakamura (AIST) for comments on the coupled oscillator model.
This work was supported by Grant-in-Aid (KAKENHI) for Scientific Research Grant B No.~JP19H01861
from Ministry of Education, Culture, Sports, Science, and Technology,
Grant-in-Aid (KAKENHI) for Transformative Research Areas B JP20H05736,
from Ministry of Education, Culture, Sports, Science, and Technology,
and JST, PRESTO Grant Number JPMJPR1992.

\appendix

\section{Memory Kernel and Noise in Coupled Oscillator Model}
\label{memory_kernel_and_noise_in_couplsed_oscillator_model}

In this appendix, we show the detailed calculations for the memory
kernel and the noise for the transient potential in the coupled
oscillator model. We derive the dynamic equation for $A(t)$ (eq~\eqref{dynamic_equation_coupled_oscillator_transient_potential_a}).
We perform the Laplace transform for eq~\eqref{dynamic_equation_coupled_oscillator_a_modified}:
\begin{equation}
 \label{dynamic_equation_coupled_oscillator_a_laplace_transform}
  \begin{split}
  s A^{*}(s) - A(0)
& =  \frac{1}{\kappa_{\text{eff}}} \sum_{j} 
   \bigg[ - 
  \frac{\kappa_{j} \omega_{j}^{2} }{\omega_{j}^{2} + s^{2}} (\theta_{j}(0) - Q(0)) 
  + \frac{\omega_{j}^{2}  s}{\omega_{j}^{2} + s^{2}} \pi_{j}(0) \\
 & \qquad + 
   \frac{\kappa_{j} \omega_{j}^{2} }{\omega_{j}^{2} + s^{2}} (s A^{*}(s) - A(0))
   + \frac{\omega_{j}^{2} s}{\omega_{j}^{2} + s^{2}} \kappa_{j} (Q^{*}(s) - A^{*}(s))
   \bigg].
  \end{split}
\end{equation}
Rearranging eq~\eqref{dynamic_equation_coupled_oscillator_a_laplace_transform}
and we have
\begin{equation}
 \label{dynamic_equation_coupled_oscillator_a_laplace_transform_arranged}
  \begin{split}
   s A^{*}(s) - A(0)
& =    \left[ 1 - \frac{1}{\kappa_{\text{eff}}} \sum_{j}   \frac{\kappa_{j} \omega_{j}^{2} }{\omega_{j}^{2} + s^{2}}\right]^{-1}
 \frac{1}{\kappa_{\text{eff}}} \sum_{j} 
   \bigg[ 
   - \frac{\omega_{j}^{2} s}{\omega_{j}^{2} + s^{2}} \kappa_{j} ( A^{*}(s) - Q^{*}(s)) \\
 & \qquad
   - 
  \frac{\kappa_{j} \omega_{j}^{2} }{\omega_{j}^{2} + s^{2}} (\theta_{j}(0) - Q(0)) 
  + \frac{\omega_{j}^{2}  s}{\omega_{j}^{2} + s^{2}} \pi_{j}(0) 
   \bigg].
  \end{split}
\end{equation}
Eq~\eqref{dynamic_equation_coupled_oscillator_a_laplace_transform_arranged} can
be rewritten as
\begin{equation}
 \label{dynamic_equation_coupled_oscillator_a_laplace_transform_modified}
   s A^{*}(s) - A(0)
 = - K^{*}(s) \kappa_{\text{eff}} ( A^{*}(s) - Q^{*}(s))
 + \xi^{*}(s) ,
\end{equation}
with $K^{*}(s)$ and $\xi^{*}(s)$ defined in eqs~\eqref{dynamic_equation_coupled_oscillator_transient_potential_memory_kernel}
and \eqref{dynamic_equation_coupled_oscillator_transient_potential_noise}.
The inverse Laplace transform of eq~\eqref{dynamic_equation_coupled_oscillator_a_laplace_transform_modified}
gives eq~\eqref{dynamic_equation_coupled_oscillator_transient_potential_a} in the main text.

We derive the fluctuation-dissipation relation (eq~\eqref{fluctuation_dissipation_relation_transient_potential_coupled_oscillator}).
The explicit expression of $\xi(t)$ in the time domain is not simple.
In the Laplace domain, from eq~\eqref{dynamic_equation_coupled_oscillator_transient_potential_noise},
we have $\langle \xi^{*}(s) \rangle_{\text{eq},0} = 0$. The inverse Laplace
transform gives
the first equation in 
eq~\eqref{fluctuation_dissipation_relation_transient_potential_coupled_oscillator}.
From the time-translational symmetry, the second order moment satisfies
$ \langle \xi(t) \xi(t') \rangle_{\text{eq},0} =  \langle \xi(t - t') \xi(0) \rangle_{\text{eq},0} $ for $t > t'$.
For $t' \le t$, from the symmetry, we have $\langle \xi(t) \xi(t') \rangle_{\text{eq},0} = \langle \xi(t' - t) \xi(0) \rangle_{\text{eq},0}$.
Then we have $\langle \xi(t) \xi(t') \rangle_{\text{eq},0} = \langle \xi(|t - t'|) \xi(0) \rangle_{\text{eq},0}$.
We calculate the Laplace transform of $\langle \xi(t) \xi(0) \rangle_{\text{eq},0} $.
By using eqs~\eqref{dynamic_equation_coupled_oscillator_transient_potential_memory_kernel}
and \eqref{gle_coupled_oscillator_noise}, we have
\begin{equation}
 \label{fluctuation_dissipation_relation_transient_potential_coupled_oscillator_laplace_transform}
 \langle \xi^{*}(s) \xi(0) \rangle_{\text{eq},0}
  =   k_{B} T
    \left[ 1 + \frac{1}{\kappa_{\text{eff}}} \sum_{j} \frac{\kappa_{j} \omega_{j}^{2} }{\omega_{j}^{2} + s^{2}} \right]^{-1} 
  \frac{1}{\kappa_{\text{eff}}^{2}} \sum_{j}
  \frac{\kappa_{j} \omega_{j}^{2}  s}{\omega_{j}^{2} + s^{2}} = k_{B} T K^{*}(s).
\end{equation}
The inverse Laplace transform of eq~\eqref{fluctuation_dissipation_relation_transient_potential_coupled_oscillator_laplace_transform} gives the second equation
in eq~\eqref{fluctuation_dissipation_relation_transient_potential_coupled_oscillator}.
Thus we have
eq~\eqref{fluctuation_dissipation_relation_transient_potential_coupled_oscillator}
in the main text.

\section{Some Relations which Involve Liouville and Projection Operators}
\label{some_relations_which_involve_liouville_and_projection_operators}

In this appendix, we show some relations which involve Liouville and
projection operators. These relations are utilized to derive the dynamic
equation for the transient potential in the main text, eq~\eqref{effective_dynamic_equation_phi}.

From the definition of the Liouville operator (eq~\eqref{liouville_operator_gamma}), 
we have $\int d\bm{\Gamma} \, \hat{f}(\bm{\Gamma}) \mathcal{L} \hat{g}(\bm{\Gamma})
= \int d\bm{\Gamma} \, [- \mathcal{L} \hat{f}(\bm{\Gamma})]\hat{g}(\bm{\Gamma})$,
$\mathcal{L} \hat{\Psi}_{\text{eq}}(\bm{\Gamma}) = 0$, and $\mathcal{L}[\hat{f}(\bm{\Gamma})\hat{g}(\bm{\Gamma})] = \hat{f}(\bm{\Gamma}) \mathcal{L} \hat{g}(\bm{\Gamma}) + [\mathcal{L} \hat{f}(\bm{\Gamma})] \hat{g}(\bm{\Gamma})$
where $\hat{f}(\bm{\Gamma}_{0})$ and $\hat{g}(\bm{\Gamma}_{0})$ are arbitrary functions of $\bm{\Gamma}_{0}$. Thus we have
the following relation:
\begin{equation}
  \label{liouville_operator_conjugate_psi}
\begin{split}
 \int d\bm{\Gamma} \, \hat{\Psi}_{\text{eq}}(\bm{\Gamma}) \hat{f}(\bm{\Gamma}) \mathcal{L} \hat{g}(\bm{\Gamma})
 & =  \int d\bm{\Gamma} \, [- \mathcal{L} (\hat{\Psi}_{\text{eq}}(\bm{\Gamma}) \hat{f}(\bm{\Gamma}))] \hat{g}(\bm{\Gamma}) \\
 & = \int d\bm{\Gamma} \, \hat{\Psi}_{\text{eq}}(\bm{\Gamma}) [- \mathcal{L} \hat{f}(\bm{\Gamma})] \hat{g}(\bm{\Gamma}) .
\end{split}
\end{equation}
If the Liouville operator is operated to the product of the delta functions and delta functional
for the coarse-grained variables, we have
\begin{equation}
 \label{time_evolution_delta_functions}
 \begin{split}
  & \mathcal{L} [\delta(\bm{Q} - \bm{Q}') \delta(\bm{P} - \bm{P}') \delta[\hat{U}(\cdot,\bm{\theta}) - \hat{U}(\cdot,\bm{\theta}')] ] \\
  & = 
  -
  \left[ (\mathcal{L} \bm{Q}) \cdot \frac{\partial }{\partial \bm{Q}'} 
  + (\mathcal{L} \bm{P}) \cdot  \frac{\partial }{\partial \bm{P}'} 
  + \int d\tilde{\bm{q}}' \, [\mathcal{L} \hat{U}(\tilde{\bm{q}}',\bm{\theta})] \frac{\delta }{\delta \hat{U}(,\tilde{\bm{q}}',\bm{\theta}')} \right] \\
  & \qquad \times [\delta(\bm{Q} - \bm{Q}') \delta(\bm{P} - \bm{P}') \delta[\hat{U}(\cdot,\bm{\theta}) - \hat{U}(\cdot,\bm{\theta}')] ]\\
  & = 
  -
  \left[ \bm{P} \cdot \bm{M}^{-1} \cdot \frac{\partial }{\partial \bm{Q}'} 
  - \frac{\partial \hat{U}(\bm{Q},\bm{\theta})}{\partial \bm{Q}} \cdot  \frac{\partial }{\partial \bm{P}'} 
  + \int d\tilde{\bm{q}}' \, [\mathcal{L} \hat{U}(\tilde{\bm{q}}',\bm{\theta})] \frac{\delta }{\delta \hat{U}(\tilde{\bm{q}}',\bm{\theta}')} \right] \\
  & \qquad \times [\delta(\bm{Q} - \bm{Q}') \delta(\bm{P} - \bm{P}') \delta[\hat{U}(\cdot,\bm{\theta}) - \hat{U}(\cdot,\bm{\theta}')] ].
 \end{split}
\end{equation}

For the projection operator $\mathcal{P}$, we have the following relation:
\begin{equation}
 \label{projection_operator_p_conjugate_psi}
\begin{split}
 & \int d\bm{\Gamma}_{0} \, \hat{\Psi}_{\text{eq}}(\bm{\Gamma}_{0}) \hat{f}(\bm{\Gamma}_{0}) \mathcal{P} \hat{g}(\bm{\Gamma}_{0}) \\
 & = \int d\bm{\Gamma}_{0} d\bm{\Gamma}_{0}' \, 
  \frac{\delta(\bm{Q}_{0} - \bm{Q}_{0}') \delta(\bm{P}_{0} - \bm{P}_{0}') \delta[\hat{U}(\cdot,\bm{\theta}_{0}) - \hat{U}(\cdot,\bm{\theta}_{0}')]}
  {\bar{\Psi}_{\text{eq}}[\bm{Q}_{0},\bm{P}_{0},\hat{U}(\cdot,\bm{\theta}_{0})]} \hat{\Psi}_{\text{eq}}(\bm{\Gamma}_{0}) \hat{f}(\bm{\Gamma}_{0}) \hat{\Psi}_{\text{eq}}(\bm{\Gamma}_{0}') \hat{g}(\bm{\Gamma}_{0}') \\
 & = \int d\bm{\Gamma}_{0} \, \hat{\Psi}_{\text{eq}}(\bm{\Gamma}_{0}) [\mathcal{P} \hat{f}(\bm{\Gamma}_{0})] \hat{g}(\bm{\Gamma}_{0}).
\end{split}
\end{equation}
From $\mathcal{P} 1 = 1$, we also have
\begin{equation}
  \int d\bm{\Gamma}_{0} \, \hat{\Psi}_{\text{eq}}(\bm{\Gamma}_{0}) \hat{f}(\bm{\Gamma}_{0}) 
  = \int d\bm{\Gamma}_{0} \, \hat{\Psi}_{\text{eq}}(\bm{\Gamma}_{0}) \mathcal{P} \hat{f}(\bm{\Gamma}_{0}) . 
\end{equation}
The projection for the product which involves the delta functions and delta functional for the coarse-grained variables becomes
\begin{equation}
 \label{projection_operator_p_delta_functions}
 \begin{split}
  & \mathcal{P} \left[ \hat{f}(\bm{\Gamma}_{0})
  \delta(\bm{Q}_{0} - \bm{Q}_{0}') \delta(\bm{P}_{0} - \bm{P}_{0}') \delta[\hat{U}(\cdot,\bm{\theta}_{0}) - \hat{U}(\cdot,\bm{\theta}_{0}')] \right] \\
  & = \frac{1}{\bar{\Psi}_{\text{eq}}[\bm{Q}_{0},\bm{P}_{0},\hat{U}(\cdot,\bm{\theta}_{0})]}
  \int d\bm{\Gamma}_{0}'' \, \hat{\Psi}_{\text{eq}}(\bm{\Gamma}_{0}'')
  \hat{f}(\bm{\Gamma}_{0}'') \delta(\bm{Q}_{0} - \bm{Q}_{0}'') \delta(\bm{P}_{0} - \bm{P}_{0}'') \\
  & \qquad \times \delta[\hat{U}(\cdot,\bm{\theta}_{0}) - \hat{U}(\cdot,\bm{\theta}_{0}'')] \delta(\bm{Q}_{0}'' - \bm{Q}_{0}') \delta(\bm{P}_{0}'' - \bm{P}_{0}') \delta[\hat{U}(\cdot,\bm{\theta}_{0}'') - \hat{U}(\cdot,\bm{\theta}_{0}')] \\
  & = \frac{  \delta(\bm{Q}_{0} - \bm{Q}_{0}') \delta(\bm{P}_{0} - \bm{P}_{0}') \delta[\hat{U}(\cdot,\bm{\theta}_{0}) - \hat{U}(\cdot,\bm{\theta}_{0}')]}{\bar{\Psi}_{\text{eq}}[\bm{Q}_{0},\bm{P}_{0},\hat{U}(\cdot,\bm{\theta}_{0})]}
  \int d\bm{\Gamma}_{0}'' \, \hat{\Psi}_{\text{eq}}(\bm{\Gamma}_{0}'') \hat{f}(\bm{\Gamma}_{0}'') \\ 
  & \qquad \times \delta(\bm{Q}_{0} - \bm{Q}_{0}'') \delta(\bm{P}_{0} - \bm{P}_{0}'') \delta[\hat{U}(\cdot,\bm{\theta}_{0}) - \hat{U}(\cdot,\bm{\theta}_{0}'')] \\
  & = [\mathcal{P} \hat{f}(\bm{\Gamma}_{0})] 
   \delta(\bm{Q}_{0} - \bm{Q}_{0}') \delta(\bm{P}_{0} - \bm{P}_{0}') \delta[\hat{U}(\cdot,\bm{\theta}_{0}) - \hat{U}(\cdot,\bm{\theta}_{0}')] ,
 \end{split}
\end{equation}
where $\bm{\Gamma}_{0}'' = [\bm{Q}_{0}'',\bm{P}_{0}'', \bm{\theta}_{0}'', \bm{\pi}_{0}'']$ is the initial position
in the phase space.
Similar relations to eqs \eqref{projection_operator_p_conjugate_psi} and
\eqref{projection_operator_p_delta_functions}
hold for $\mathcal{Q}$:
\begin{equation}
 \label{projection_operator_q_conjugate_psi}
  \int d\bm{\Gamma} \, \hat{\Psi}_{\text{eq}}(\bm{\Gamma}) \hat{f}(\bm{\Gamma}) \mathcal{Q}(\bm{\Gamma}) \hat{g}(\bm{\Gamma})
   = \int d\bm{\Gamma} \, \hat{\Psi}_{\text{eq}}(\bm{\Gamma}) [\mathcal{Q} \hat{f}(\bm{\Gamma})] \hat{g}(\bm{\Gamma}),
\end{equation}
\begin{equation}
 \label{projection_operator_q_delta_functions}
  \begin{split}
   & \mathcal{Q} 
   \left[ \hat{f}(\bm{\Gamma}_{0})
    \delta(\bm{Q}_{0} - \bm{Q}_{0}') \delta(\bm{P}_{0} - \bm{P}_{0}') \delta[\hat{U}(\cdot,\bm{\theta}_{0}) - \hat{U}(\cdot,\bm{\theta}_{0}')]
   \right] \\
   & = [\mathcal{Q} \hat{f}(\bm{\Gamma}_{0})] 
    \delta(\bm{Q}_{0} - \bm{Q}_{0}') \delta(\bm{P}_{0} - \bm{P}_{0}') \delta[\hat{U}(\cdot,\bm{\theta}_{0}) - \hat{U}(\cdot,\bm{\theta}_{0}')].
  \end{split}
\end{equation}

For the combination of the Liouville operator and the projection operator $\mathcal{Q}$, we have
\begin{equation}
 \label{projected_time_evolution_operator_shift}
 \mathcal{Q} e^{t \mathcal{L} \mathcal{Q}}
  = \mathcal{Q} \sum_{n = 0}^{\infty} \frac{1}{n!} (t \mathcal{L} \mathcal{Q})^{n}
  =\sum_{n = 0}^{\infty} \frac{1}{n!} (t  \mathcal{Q} \mathcal{L})^{n} \mathcal{Q}
  = e^{t \mathcal{Q} \mathcal{L}} \mathcal{Q} ,
\end{equation}
and
\begin{equation}
 \label{adjoint_projected_time_evolution_operator}
\begin{split}
 & \int d\bm{\Gamma}_{0} \, \hat{\Psi}_{\text{eq}}(\bm{\Gamma}_{0}) \hat{f}(\bm{\Gamma}_{0}) e^{t \mathcal{Q} \mathcal{L}} \hat{g}(\bm{\Gamma}_{0})
  =  \sum_{n = 0}^{\infty} \frac{1}{n!} \int d\bm{\Gamma}_{0} \, \hat{\Psi}_{\text{eq}}(\bm{\Gamma}_{0}) \hat{f}(\bm{\Gamma}_{0}) 
  (t \mathcal{Q} \mathcal{L})^{n} \hat{g}(\bm{\Gamma}_{0}) \\
 & = \int d\bm{\Gamma}_{0} \, \hat{\Psi}_{\text{eq}}(\bm{\Gamma}_{0}) \sum_{n = 0}^{\infty} \frac{1}{n!}  [(t \mathcal{Q} \mathcal{L})^{n} \hat{f}(\bm{\Gamma}_{0})]
  \hat{g}(\bm{\Gamma}_{0})
  = 
 \int d\bm{\Gamma}_{0} \, \hat{\Psi}_{\text{eq}}(\bm{\Gamma}_{0}) [e^{t \mathcal{L} \mathcal{Q}} \hat{f}(\bm{\Gamma}_{0})] \hat{g}(\bm{\Gamma}_{0}).
\end{split}
\end{equation}

We calculate the expression 
$\mathcal{P} \mathcal{L} \mathcal{Q} \hat{\Xi}(\bm{\Gamma}_{0},\tilde{\bm{q}},t - t')$
in the damping term (eq~\eqref{transient_potential_damping_term_with_fluctuating_term}).
We first calculate $\mathcal{P} \mathcal{L} \mathcal{Q} \hat{f}(\bm{\Gamma}_{0})$ 
(with $\hat{f}(\bm{\Gamma}_{0})$ being an arbitrary function of $\bm{\Gamma}_{0}$)
and then we substitute
$\hat{\Xi}(\bm{\Gamma}_{0},\tilde{\bm{q}},t - t')$ for $\hat{f}(\bm{\Gamma}_{0})$.
From eqs~\eqref{liouville_operator_conjugate_psi}
and \eqref{projection_operator_q_conjugate_psi}, we have
\begin{equation}
 \label{doubly_projected_time_evolution}
 \begin{split}
  \mathcal{P}\mathcal{L} \mathcal{Q} \hat{f}(\bm{\Gamma}_{0})
  & = \frac{1}{\bar{\Psi}_{\text{eq}}[\bm{Q}_{0},\bm{P}_{0},\hat{U}(\cdot,\bm{\theta}_{0})]}
  \int d\bm{\Gamma}_{0}' \, \hat{\Psi}_{\text{eq}}(\bm{\Gamma}_{0}')
    \delta(\bm{Q}_{0} - \bm{Q}_{0}') \delta(\bm{P}_{0} - \bm{P}_{0}') \\
  & \qquad \times \delta[\hat{U}(\cdot,\bm{\theta}_{0}) - \hat{U}(\cdot,\bm{\theta}_{0}')] \mathcal{L}' \mathcal{Q}' \hat{f}(\bm{\Gamma}_{0}') \\
  & = - \frac{1}{\bar{\Psi}_{\text{eq}}[\bm{Q}_{0},\bm{P}_{0},\hat{U}(\cdot,\bm{\theta}_{0})]}
  \int d\bm{\Gamma}_{0}' \,  \hat{\Psi}_{\text{eq}}(\bm{\Gamma}_{0}') [\mathcal{Q}' \hat{f}(\bm{\Gamma}_{0}')]  \\
  & \qquad \times \mathcal{L}' \left[ \delta(\bm{Q}_{0} - \bm{Q}_{0}') \delta(\bm{P}_{0} - \bm{P}_{0}') \delta[\hat{U}(\cdot,\bm{\theta}_{0}) - \hat{U}(\cdot,\bm{\theta}_{0}')]\right] \\
  & = - \frac{1}{\bar{\Psi}_{\text{eq}}[\bm{Q}_{0},\bm{P}_{0},\hat{U}(\cdot,\bm{\theta}_{0})]}
  \int d\bm{\Gamma}_{0}' \,   \hat{f}(\bm{\Gamma}_{0}') \hat{\Psi}_{\text{eq}}(\bm{\Gamma}_{0}') \\
  & \qquad \times \mathcal{Q}'
   \mathcal{L}' \left[ \delta(\bm{Q}_{0} - \bm{Q}_{0}') \delta(\bm{P}_{0} - \bm{P}_{0}') \delta[\hat{U}(\cdot,\bm{\theta}_{0}) - \hat{U}(\cdot,\bm{\theta}_{0}')]\right] .
 \end{split}
\end{equation}
Here, $\mathcal{L}',\mathcal{Q}'$ are the Liouville operator and the projection operator for $\bm{\Gamma}'$.
We find that operators $\mathcal{Q}'$ and $\mathcal{L}'$ are operated to the product of delta functions in the integrand of eq~\eqref{doubly_projected_time_evolution}.
We can utilize the relations \eqref{time_evolution_delta_functions} and \eqref{projection_operator_q_delta_functions}
to simplify eq~\eqref{doubly_projected_time_evolution}.
\begin{equation}
 \label{doubly_projected_time_evolution_integrand}
 \begin{split}
  & \mathcal{Q}'
   \mathcal{L}' \left[ \delta(\bm{Q}_{0} - \bm{Q}_{0}') \delta(\bm{P}_{0} - \bm{P}_{0}') \delta[\hat{U}(\cdot,\bm{\theta}_{0}) - \hat{U}(\cdot,\bm{\theta}_{0}')]\right] \\
  & =   - \mathcal{Q}'
  \left[ \bm{P}_{0}' \cdot \bm{M}^{-1} \cdot \frac{\partial }{\partial \bm{Q}_{0}} 
  - \frac{\partial \hat{U}(\bm{Q}_{0}',\bm{\theta}_{0}')}{\partial \bm{Q}_{0}'} \cdot  \frac{\partial }{\partial \bm{P}_{0}} 
  + \int d\tilde{\bm{q}}' \, (\mathcal{L}' \hat{U}(\tilde{\bm{q}}',\bm{\theta}_{0}')) \frac{\delta }{\delta \hat{U}(\tilde{\bm{q}}',\bm{\theta}_{0})} \right] \\
  & \qquad \times \left[ \delta(\bm{Q}_{0} - \bm{Q}_{0}') \delta(\bm{P}_{0} - \bm{P}_{0}') \delta[\hat{U}(\cdot,\bm{\theta}_{0}) - \hat{U}(\cdot,\bm{\theta}_{0}')]\right] \\
  & =  - \mathcal{Q}'
   \int d\tilde{\bm{q}}' \, (\mathcal{L}' \hat{U}(\tilde{\bm{q}}',\bm{\theta}_{0}')) \frac{\delta }{\delta \hat{U}(\tilde{\bm{q}}',\bm{\theta}_{0})} 
\left[ \delta(\bm{Q}_{0} - \bm{Q}_{0}') \delta(\bm{P}_{0} - \bm{P}_{0}') \delta[\hat{U}(\cdot,\bm{\theta}_{0}) - \hat{U}(\cdot,\bm{\theta}_{0}')]\right] \\
  & =  - 
   \int d\tilde{\bm{q}}' \, \frac{\delta }{\delta \hat{U}(\tilde{\bm{q}}',\bm{\theta}_{0})} 
  \mathcal{Q}' \left[[\mathcal{L}' \hat{U}(\tilde{\bm{q}}',\bm{\theta}_{0}')]
\delta(\bm{Q}_{0} - \bm{Q}_{0}') \delta(\bm{P}_{0} - \bm{P}_{0}') \delta[\hat{U}(\cdot,\bm{\theta}_{0}) - \hat{U}(\cdot,\bm{\theta}_{0}')]
  \right]  \\
  & =  - 
   \int d\tilde{\bm{q}}' \, \frac{\delta }{\delta \hat{U}(\tilde{\bm{q}}',\bm{\theta}_{0})} 
\left[[ \mathcal{Q}' \mathcal{L}' \hat{U}(\tilde{\bm{q}}',\bm{\theta}_{0}') ]
\delta(\bm{Q}_{0} - \bm{Q}_{0}') \delta(\bm{P}_{0} - \bm{P}_{0}') \delta[\hat{U}(\cdot,\bm{\theta}_{0}) - \hat{U}(\cdot,\bm{\theta}_{0}')] \right] .
 \end{split}
\end{equation}
From eqs~\eqref{doubly_projected_time_evolution_integrand} and \eqref{doubly_projected_time_evolution},
we have
\begin{equation}
 \label{doubly_projected_time_evolution_modified}
 \begin{split}
  \mathcal{P}\mathcal{L} \mathcal{Q} \hat{f}(\bm{\Gamma}_{0})
  & = \frac{1}{\bar{\Psi}_{\text{eq}}[\bm{Q}_{0},\bm{P}_{0},\hat{U}(\cdot,\bm{\theta}_{0})]}
  \int d\tilde{\bm{q}}' \, \frac{\delta}{\delta \hat{U}(\tilde{\bm{q}}',\bm{\theta}_{0})} \int d\bm{\Gamma}_{0}' \,   \hat{f}(\bm{\Gamma}_{0}') \hat{\Psi}_{\text{eq}}(\bm{\Gamma}_{0}') \\
  & \qquad \times 
  \left[ [\mathcal{Q}' \mathcal{L}' \hat{U}(\tilde{\bm{q}}',\bm{\theta}_{0}')]
  \delta(\bm{Q}_{0} - \bm{Q}_{0}') \delta(\bm{P}_{0} - \bm{P}_{0}') \delta[\hat{U}(\cdot,\bm{\theta}_{0}) - \hat{U}(\cdot,\bm{\theta}_{0}')]  \right] \\
  & = \frac{1}{\bar{\Psi}_{\text{eq}}[\bm{Q}_{0},\bm{P}_{0},\hat{U}(\cdot,\bm{\theta}_{0})]}
  \int d\tilde{\bm{q}}' 
  \, \frac{\delta}{\delta \hat{U}(\tilde{\bm{q}}',\bm{\theta}_{0})} \\
  & \qquad \times \left[ \bar{\Psi}_{\text{eq}}[\bm{Q}_{0},\bm{P}_{0},\hat{U}(\cdot,\bm{\theta}_{0})]
  \, \mathcal{P}
  \left[ \hat{f}(\bm{\Gamma}_{0})  (\mathcal{Q} \mathcal{L} \hat{U}(\tilde{\bm{q}}',\bm{\theta}_{0})) \right]
   \right] .
 \end{split}
\end{equation}
Finally, by substituting $\hat{f}(\bm{\Gamma}_{0}) = \hat{\Xi}(\bm{\Gamma}_{0},\tilde{\bm{q}},t - t')$ into
eq~\eqref{doubly_projected_time_evolution_modified} and
using $\hat{\Xi}(\bm{\Gamma}_{0},\tilde{\bm{q}},0)
= \mathcal{Q} \mathcal{L} \hat{U}(\tilde{\bm{q}},\bm{\theta}_{0})$, we have
eq~\eqref{transient_potential_damping_term_modified} in the main text.

We calculate the equilibrium average of the second order moment of the fluctuating term
by utilizing eqs~\eqref{projected_time_evolution_operator_shift} and \eqref{adjoint_projected_time_evolution_operator}.
For $t > t'$, we have
\begin{equation}
 \label{transient_potential_fluctuating_term_average_second_order_moment_modified}
 \begin{split}
  \left\langle \hat{\Xi}(\bm{\Gamma}_{0},\tilde{\bm{q}},t)
  \hat{\Xi}(\bm{\Gamma}_{0},\tilde{\bm{q}}',t') 
  \right\rangle_{\text{eq},0}
  & = \int d\bm{\Gamma}_{0} \, \hat{\Psi}_{\text{eq}}(\bm{\Gamma}_{0})
  [\mathcal{Q} e^{t \mathcal{L} \mathcal{Q}} \mathcal{L} \hat{U}(\tilde{\bm{q}},\bm{\theta}_{0}) ]
  [\mathcal{Q} e^{t' \mathcal{L} \mathcal{Q}} \mathcal{L} \hat{U}(\tilde{\bm{q}}',\bm{\theta}_{0}) ] \\
  & = \int d\bm{\Gamma}_{0} \, \hat{\Psi}_{\text{eq}}(\bm{\Gamma}_{0})
  [\mathcal{Q} e^{t \mathcal{L} \mathcal{Q}} \mathcal{L} \hat{U}(\tilde{\bm{q}},\bm{\theta}_{0}) ]
  [e^{t' \mathcal{Q}\mathcal{L} }\mathcal{Q} \mathcal{L} \hat{U}(\tilde{\bm{q}}',\bm{\theta}_{0}) ] \\
  & = \int d\bm{\Gamma}_{0} \, \hat{\Psi}_{\text{eq}}(\bm{\Gamma}_{0})
  [e^{- t'\mathcal{L}  \mathcal{Q}} \mathcal{Q}  e^{t \mathcal{L} \mathcal{Q}}  \mathcal{L}\hat{U}(\tilde{\bm{q}},\bm{\theta}_{0}) ]
  \mathcal{Q}\mathcal{L}\hat{U}(\tilde{\bm{q}}',\bm{\theta}_{0}) .
 \end{split}
\end{equation}
Here we utilize the following relations for operators: $\mathcal{Q} = \mathcal{Q}^{2}$ and $\mathcal{Q} e^{t \mathcal{L} \mathcal{Q}} \mathcal{Q} = \mathcal{Q} e^{t \mathcal{L} \mathcal{Q}}$.
Then eq~\eqref{transient_potential_fluctuating_term_average_second_order_moment_modified}
can be further modified as
\begin{equation}
 \label{transient_potential_fluctuating_term_average_second_order_moment_modified2}
 \begin{split}
  \left\langle \hat{\Xi}(\bm{\Gamma}_{0},\tilde{\bm{q}},t)
  \hat{\Xi}(\bm{\Gamma}_{0},\tilde{\bm{q}}',t') 
  \right\rangle_{\text{eq},0}
  & = \int d\bm{\Gamma}_{0} \, \hat{\Psi}_{\text{eq}}(\bm{\Gamma}_{0})
  [e^{(t - t') \mathcal{L} \mathcal{Q}} \mathcal{Q} \mathcal{L}\hat{U}(\tilde{\bm{q}},\bm{\theta}_{0}) ]
  \mathcal{Q}^{2} \mathcal{L}\hat{U}(\tilde{\bm{q}}',\bm{\theta}_{0}) \\
  & = \int d\bm{\Gamma}_{0} \, \hat{\Psi}_{\text{eq}}(\bm{\Gamma}_{0})
  [\mathcal{Q} e^{(t - t') \mathcal{L} \mathcal{Q}}  \mathcal{L}\hat{U}(\tilde{\bm{q}},\bm{\theta}_{0}) ]
  \mathcal{Q} \mathcal{L}\hat{U}(\tilde{\bm{q}}',\bm{\theta}_{0}) \\
  & = \int d\bm{\Gamma}_{0} \, \hat{\Psi}_{\text{eq}}(\bm{\Gamma}_{0}) \,
  \hat{\Xi}(\tilde{\bm{q}},\bm{\Gamma}_{0},t - t') \hat{\Xi}(\tilde{\bm{q}},\bm{\Gamma}_{0},0) ,
 \end{split}
\end{equation}
where we have utilized eq~\eqref{projection_operator_q_conjugate_psi}.
Eq~\eqref{transient_potential_fluctuating_term_average_second_order_moment_modified2}
together with eq~\eqref{memory_kernel_phi}
gives
\begin{equation}
 \label{transient_potential_fluctuating_term_average_second_order_moment}
\begin{split}
 \left\langle \hat{\Xi}(\bm{\Gamma}_{0},\tilde{\bm{q}},t)
   \hat{\Xi}(\bm{\Gamma}_{0},\tilde{\bm{q}}',t')
  \right\rangle_{\text{eq},0}
 & = k_{B} T \int d\bm{\Gamma}_{0} \, \hat{\Psi}_{\text{eq}}(\bm{\Gamma}_{0})
 \frac{1}{k_{B} T} \mathcal{P} 
\left[ \hat{\Xi}(\bm{\Gamma}_{0},\tilde{\bm{q}},t)
   \hat{\Xi}(\bm{\Gamma}_{0},\tilde{\bm{q}}',t') \right] \\
 & = k_{B} T
  \left\langle 
  \bar{K}[\bm{Q}_{0},\bm{P}_{0},\hat{U}(\cdot,\bm{\theta}_{0}),\tilde{\bm{q}},\tilde{\bm{q}}',t - t'] \right\rangle_{\text{eq},0} .
\end{split}
\end{equation}

For $t < t'$, we utilize the time-reversal transform.
Under the time-reversal transform, the equilibrium probability distribution is not changed but the time-evolution of the system is reversed.
There is a reverse path in the phase space for a given path, and the probabilities to find a path and its reverse path are the same.
Then the correlation function \eqref{transient_potential_fluctuating_term_average_second_order_moment} should be symmetric under the transform $t \to - t$. This consideration gives
eq~\eqref{transient_potential_fluctuating_term_average_second_order_moment_symmetrized}
in the main text.

\section{Calculation of Kullback-Leibler Divergence}
\label{calculation_of_kullback_leibler_divergence}

In this appendix, we show the details of the calculation for the
approximate form of the Kullback-Leibler divergence (eq~\eqref{kullback_leibler_divergence_approx}).
By utilizing the approximate form for the transient potential, eq~\eqref{transient_potential_harmonic_approximation},
we have the following explicit expression for the equilibrium probability distribution:
\begin{equation}
 \label{equilibrium_distribution_harmonic_approx}
\begin{split}
  \Phi_{\text{eq}}(\tilde{\bm{q}})
 & \approx \mathcal{N} \exp
 \left[ - \frac{1}{k_{B} T}
 \left[ \Phi(\bm{Q}) - \bm{F} \cdot \Delta \tilde{\bm{q}}
  + \frac{1}{2} \Delta \tilde{\bm{q}} \cdot \bm{C} \cdot \Delta \tilde{\bm{q}}
  +  \frac{1}{2}\Delta \tilde{\bm{q}} \cdot \bm{C}_{\text{trap}} \cdot \Delta \tilde{\bm{q}}
 \right]
 \right] \\
 & \propto \exp
 \left[ - \frac{1}{2 k_{B} T}
 (\Delta \tilde{\bm{q}} - \bm{\rho}) \cdot 
 (\bm{C}_{\text{trap}} + \bm{C}) \cdot
 (\Delta \tilde{\bm{q}} - \bm{\rho})
 \right] ,
\end{split}
\end{equation}
where $\mathcal{N}$ is the normalization factor and $\bm{\rho} \equiv (\bm{C}_{\text{trap}} + \bm{C})^{-1} \cdot \bm{F}$.
Since eq~\eqref{equilibrium_distribution_harmonic_approx} is a Gaussian distribution,
the normalization factor can be calculated straightforwardly.
The normalized distribution function is
\begin{equation}
 \label{equilibrium_distribution_harmonic_approx_explicit}
  \Phi_{\text{eq}}(\tilde{\bm{q}})
  \approx \sqrt{\det \left( \frac{\bm{C}_{\text{trap}} + \bm{C}}{2 \pi k_{B} T}\right)} \exp
 \left[ - \frac{1}{2 k_{B} T}
 (\Delta \tilde{\bm{q}} - \bm{\rho}) \cdot 
 (\bm{C}_{\text{trap}} + \bm{C}) \cdot
 (\Delta \tilde{\bm{q}} - \bm{\rho})
 \right] .
\end{equation}
In the same way, we have the equilibrium probability distribution
for the hypothetical transient potential:
\begin{equation}
 \label{hypothetical_equilibrium_distribution_harmonic_approx_explicit}
  \Phi_{\text{eq}}(\tilde{\bm{q}})
  \approx \sqrt{\det \left( \frac{\bm{C}_{\text{trap}} + \check{\bm{C}}}{2 \pi k_{B} T}\right)} \exp
 \left[ - \frac{1}{2 k_{B} T}
 (\Delta \tilde{\bm{q}} - \check{\bm{\rho}}) \cdot 
 (\bm{C}_{\text{trap}} + \check{\bm{C}}) \cdot
 (\Delta \tilde{\bm{q}} - \check{\bm{\rho}})
 \right] ,
\end{equation}
with $\check{\bm{\rho}} \equiv (\bm{C}_{\text{trap}} + \check{\bm{C}})^{-1} \cdot \check{\bm{F}}$.

We calculate the Kullback-Leibler divergence with these approximate
equilibrium probability distributions.
By substituting eqs~\eqref{equilibrium_distribution_harmonic_approx_explicit}
and \eqref{hypothetical_equilibrium_distribution_harmonic_approx_explicit} into
eq~\eqref{kullback_leibler_divergence}, we have
\begin{equation}
 \label{kullback_leibler_divergence_with_equilibrium_distribution_approx}
  \begin{split}
  \mathcal{K}(\tilde{\bm{a}})
   & \approx \int d\tilde{\bm{q}} \, \check{\Phi}_{\text{eq}}(\tilde{\bm{q}},\tilde{\bm{a}})
   \Bigg[
   \frac{1}{2} \ln \frac{\det (\bm{C}_{\text{trap}} + \check{\bm{C}})}{\det (\bm{C}_{\text{trap}} + \bm{C})}
   - \frac{1}{2 k_{B} T}  (\Delta \tilde{\bm{q}} - \check{\bm{\rho}}) \cdot 
 (\bm{C}_{\text{trap}} + \check{\bm{C}}) \cdot
 (\Delta \tilde{\bm{q}} - \check{\bm{\rho}}) \\
   & \qquad + \frac{1}{2 k_{B} T}  (\Delta \tilde{\bm{q}} - \bm{\rho}) \cdot 
 (\bm{C}_{\text{trap}} + \bm{C}) \cdot
 (\Delta \tilde{\bm{q}} - \bm{\rho})
   \Bigg] \\
   & = 
   \frac{1}{2} \ln \det [(\bm{C}_{\text{trap}} + \bm{C})^{-1} \cdot (\bm{C}_{\text{trap}} + \check{\bm{C}})]
   - \frac{1}{2 k_{B} T} \mathop{\mathrm{tr}} [(\bm{C}_{\text{trap}} + \check{\bm{C}}) \cdot (\bm{C}_{\text{trap}} + \check{\bm{C}})^{-1}]  \\
   & \qquad + \frac{1}{2 k_{B} T} 
   \left[ \mathop{\mathrm{tr}} [(\bm{C}_{\text{trap}} + \bm{C}) \cdot (\bm{C}_{\text{trap}} + \check{\bm{C}})^{-1}]
   +  (\check{\bm{\rho}} - \bm{\rho}) \cdot
 (\bm{C}_{\text{trap}} + \bm{C}) \cdot (\check{\bm{\rho}} - \bm{\rho}) \right] .
  \end{split}
\end{equation}
Eq~\eqref{kullback_leibler_divergence_with_equilibrium_distribution_approx} can
be simplified by introducing $\bm{G} \equiv (\bm{C}_{\text{trap}} + \bm{C})^{-1}$, $\Delta \check{\bm{C}} \equiv \check{\bm{C}} - \bm{C}$,
and $\Delta \check{\bm{\rho}} \equiv \check{\bm{\rho}} - \bm{\rho}$:
\begin{equation}
 \label{kullback_leibler_divergence_with_equilibrium_distribution_approx_modified2}
  \mathcal{K}(\tilde{\bm{a}})
   \approx \frac{1}{2} \ln \det (\bm{1} + \bm{G} \cdot \Delta \check{\bm{C}})
   - \frac{1}{2 k_{B} T} \mathop{\mathrm{tr}} [\Delta \check{\bm{C}} \cdot (\bm{1} + \bm{G} \cdot \Delta \check{\bm{C}})^{-1} \cdot \bm{G}]
 + \frac{1}{2 k_{B} T} \Delta \check{\bm{\rho}} \cdot \bm{G}^{-1} \Delta \check{\bm{\rho}} . 
\end{equation}
If the hypothetical transient potential is relatively close to the 
original transient potential, we expect that
$\Delta \check{\bm{C}}$ is small. Therefore, we expand terms into the power series of $\Delta \check{\bm{C}}$ and
retain only the leading order terms.
For a second order tensor $\bm{B}$ and a small parameter $\varepsilon$, $\det (\bm{1} + \varepsilon \bm{B})$ and
$(\bm{1} + \varepsilon \bm{B})^{-1}$ can be expanded as
$ \det (\bm{1} + \varepsilon \bm{B}) 
 = \bm{1} + \varepsilon \mathop{\mathrm{tr}} \bm{B}
 + (\varepsilon^{2} / 2) [(\mathop{\mathrm{tr}} \bm{B})^{2}
 + \mathop{\mathrm{tr}} \bm{B}^{2}] + O(\varepsilon^{3}) $ and
$(\bm{1} + \varepsilon \bm{B})^{-1} 
 = \bm{1} - \varepsilon \bm{B} + O(\varepsilon^{2})$, respectively.
With these expansion forms, eq~\eqref{kullback_leibler_divergence_with_equilibrium_distribution_approx_modified2}
can be approximated as follows, upto the second order in $\Delta \check{\bm{C}}$:
\begin{equation}
 \label{kullback_leibler_divergence_with_equilibrium_distribution_approx_modified3}
  \mathcal{K}(\tilde{\bm{a}})
   \approx \frac{1}{4} \mathop{\mathrm{tr}} (\bm{G} \cdot \Delta \check{\bm{C}})^{2}
 + \frac{1}{2 k_{B} T} \Delta \check{\bm{\rho}} \cdot \bm{G}^{-1} \Delta \check{\bm{\rho}} . 
\end{equation}
In addition, $\Delta \check{\bm{\rho}}$ can be approximated as
\begin{equation}
 \label{delta_rho_approx}
 \Delta \check{\bm{\rho}} =
  (\bm{1} + \bm{G} \cdot \Delta\check{\bm{C}})^{-1} \cdot \bm{G} \cdot \check{\bm{F}} - \bm{G} \cdot \bm{F}
  \approx \bm{G} \cdot [\Delta \check{\bm{F}}
  - \Delta\check{\bm{C}}\cdot \bm{G} \cdot \bm{F}], 
\end{equation}
where $\Delta \check{\bm{F}} \equiv \check{\bm{F}} - \bm{F}$.
By substituting eq~\eqref{delta_rho_approx} into eq~\eqref{kullback_leibler_divergence_with_equilibrium_distribution_approx_modified3},
we have eq~\eqref{kullback_leibler_divergence_approx} in the main text.

\bibliographystyle{apsrev4-1}
\bibliography{transient_potential_projection_operator.bib}


\end{document}